\documentclass[conference]{IEEEtran}
\IEEEoverridecommandlockouts

\usepackage{cite}                           %应用文献
\usepackage{amsmath,amssymb,amsfonts}       %数学符号
\usepackage{algorithmic}                    %算法表框
\usepackage{graphicx}                       %插入图像              
\usepackage{xcolor,color}                   %字体颜色
\usepackage{pifont}                         %勾勾叉叉
\usepackage{threeparttable}                 %三线表
\usepackage{multirow}                       %跨行
\usepackage{tabularx}                       %表格间距
\usepackage{subfigure}                      %多个子图
\usepackage{booktabs}                       %表格空间
\usepackage{hyperref}                       %超链接
\usepackage[switch]{lineno}                 %双栏行数
\usepackage{ulem}                           %删除线

\def\BibTeX{{\rm B\kern-.05em{\sc i\kern-.025em b}\kern-.08em
    T\kern-.1667em\lower.7ex\hbox{E}\kern-.125emX}}
\begin{document}

\title{Traceable AI-driven Avatars Using Multi-factors of Physical World and Metaverse \\

\thanks{$^*$ Means the corresponding author (youliangtian@163.com).}

}

\author{\IEEEauthorblockN{Kedi Yang, Zhenyong Zhang, Youliang Tian$^*$}
\IEEEauthorblockA{
\textit{State Key Laboratory of Public Big Data, College of Computer Science and Technology Guizhou University}\\
\textit{Guiyang, China} \\
\textit{kdyang.gz@gmail.com, zyzhangnew@gmail.com, youliangtian@163.com}}

}

\maketitle

\begin{abstract}
Metaverse allows users to delegate their AI models to an AI engine, which builds corresponding AI-driven avatars to provide immersive experience for other users. Since current authentication methods mainly focus on human-driven avatars and ignore the traceability of AI-driven avatars, attackers may delegate the AI models of a target user to an AI proxy program to perform impersonation attacks without worrying about being detected.

In this paper, we propose an authentication method using multi-factors to guarantee the traceability of AI-driven avatars. Firstly, we construct a user's identity model combining the manipulator's iris feature and the AI proxy's public key to ensure that an AI-driven avatar is associated with its original manipulator. Secondly, we propose a chameleon proxy signature scheme that supports the original manipulator to delegate his/her signing ability to an AI proxy. Finally, we design three authentication protocols for avatars based on the identity model and the chameleon proxy signature to guarantee the virtual-to-physical traceability including both the human-driven and AI-driven avatars.

Security analysis shows that the proposed signature scheme is unforgeability and the authentication method is able to defend against false accusation. Extensive evaluations show that the designed authentication protocols complete user login, avatar delegation, mutual authentication, and avatar tracing in about 1s,  meeting the actual application needs and helping to mitigate impersonation attacks by AI-driven avatars.

\end{abstract}


%\linenumbers

\section{Introduction}
Metaverse integrates a variety of technologies, such as artificial intelligence (AI) and virtual reality (VR), to build an immersive environment for human users and AI devices \cite{Wang2022Surv}, where people communicate and collaborate with various entities through their own digital avatars\cite{Wang2023Metaverse}. 

An avatar is a visual representation of a driving entity, which can be a human manipulator or an AI proxy engine. For a human manipulator, he/she controls wearable devices and submits the real-time information of body activities to interaction partner \cite{Lilija2021Move}, who renders it as a digital appearance called  \textit{human-driven avatar}. Leveraging AI, users are able to deploy their AI models such as face model and speech model to an AI engine \cite{Fiore2024AITeacher}, which generates a digital appearance similar to the original user called \textit{AI-driven avatar}. The details of generating human-driven and AI-driven avatars are given in the background at \ref{sec:Avatars}.

\begin{figure}[htbp]
    \centering
     \subfigure[The process of delegating an avatar]{
     		\label{fig:Attack_1}    \includegraphics[width=0.47\textwidth]{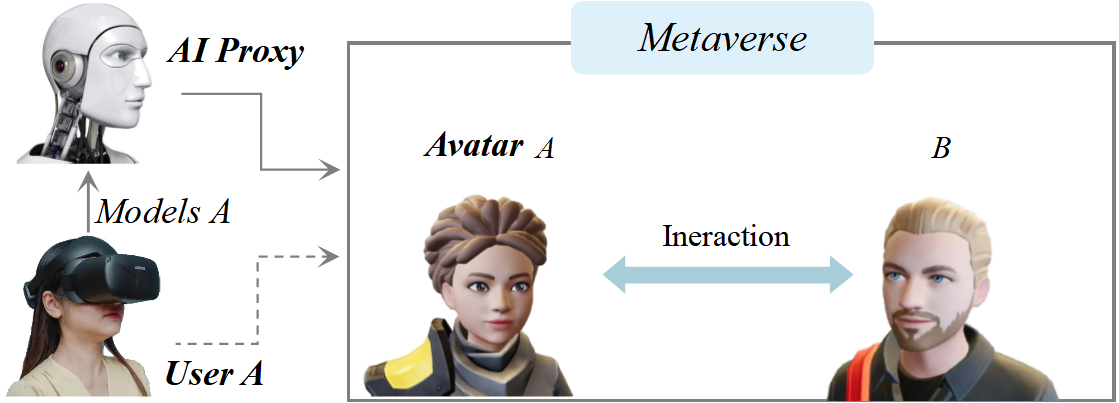}    }    \quad
    \subfigure[The  process of impersonation attack ]{
    		\label{fig:Attack_2}    \includegraphics[width=0.47\textwidth]{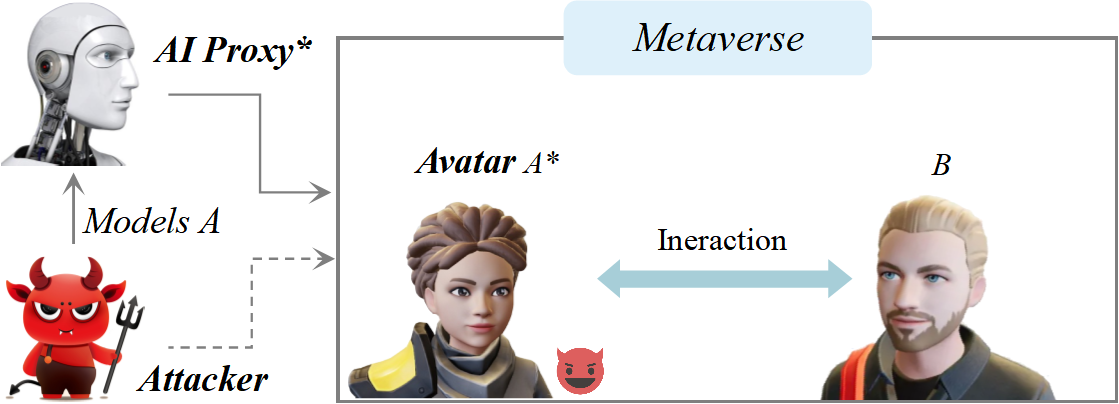}    }    \quad
    \caption{The transition process from human-driven avatar to AI-driven avatar.}
\end{figure}

A prototype of AI-driven avatar is the ``Brother Dong''\footnote{\href{https://jdcorporateblog.com/jd-com-debuts-ai-digital-representative-of-founder-richard-liu-during-livestream-drawing-20-million-views-in-under-one-hour/}{https://jdcorporateblog.com}}. In April 2024, JD.com, a large Chinese e-commerce platform, used the ChatRhino Large Language Model (LLM) to generate the AI-driven avatar Brother Dong of its founder in multiple live streaming rooms. The AI-driven avatar actively interacts with consumers and accurately mimics the founder’s expressions, voices, and hand gestures while speaking. With the further development of AI, ordinary users are able to delegate their AI models to AI proxies \cite{Yoshida2024UserAI}. For example, staffs in a company delegate their AI models to an AI proxy to generate corresponding AI-driven avatars \cite{Manfredi2023AIEdu}, which not only improve the work efficiency but also provide an immersive experience for other users. The process of delegation is shown as Fig.\ref{fig:Attack_1}. 

\subsection{Problem Statement and Security Goals}
Though the AI-driven avatars provide users with various conveniences beyond reality \cite{Wang2022Edu,Jiang2023Game, Tsai2023medicine}, there is an impersonation attack that attackers leverage a user's appearance to generate AI-driven avatars and carry out deception \cite{Soliman2023DigtHuman}. As a company in Hong Kong reported that it was deceived more than HK\$200m (£20m) because an employee received a deepfake video conference call\footnote{\href{https://www.theguardian.com/world/2024/feb/05/hong-kong-company-deepfake-video-conference-call-scam}{https://www.theguardian.com}}. During the conference, the deepfake digital appearance impersonates the senior officer of the company, who looked like the true one, and deceives the employee into transferring funds to a designated bank account. 

\textbf{Impersonation attack.} The process of impersonation attack by AI-driven avatar can be described as Fig.\ref{fig:Attack_2}.  In this process, the attacker delegates the AI models of the original user $A$ to an AI proxy, which generates the digital appearance similar with the true one to deceive other users. 

AI-driven avatars are usually generated by specific users to complete specific tasks \cite{Khampuong2023ConAgent}. For this reason, we hope to combine user's biometric features such as iris to build an identity model for AI-driven avatars, ensuring an AI-driven avatar is associated with its original manipulator. Based on this identity model, we want to design a traceable authentication method suitable for AI-driven avatars, by which victims are able to track down the original manipulator. Of course, the tracing method should ensure that legitimate users will not be falsely accused. Thus, we specifically consider these two security goals as following: 

\textbf{Virtual-to-Physical Traceability.} It refers to that a virtual avatar can be trace back to its original physical manipulator, including both human-driven and AI-driven avatars. To achieve this goal, before entering metaverse, human users need to submit their physical identity and metaverse public key to an identity provider (IdP), who serves as a regulatory organization to generate corresponding metaverse identity token (MIT) for users. With this token, a user can generate human-driven and AI-driven avatars, where the AI-driven avatar incorporates the physical feature of original manipulator.

\textbf{Defending against False Accusation.} It refers to that malicious reporters are unable to forge identity parameters to frame target user. In metaverse interaction, a malicious reporter may interact with the human-driven and AI-driven avatars of a target user to collect the user's identity parameters. Based on these parameters, the reporter attempts to forge a parameter such that it could be verified by the regulator, thereby falsely accusing the target user. For the false accusation, we require that even if an attacker is able to extensively interact with the avatars of a target user, the attacker fails to forge the corresponding identity parameters, thus defending against false accusation.

\subsection{Related Works and Their Limitations}

Although some related works have focused on the issues of avatar authentication and tracing, there are three limitations to establishing the traceable AI-driven avatars: (i) There is a lack of a user identity model that is compatible with both human users and AI proxies; (ii) There is a lack of a signature scheme that can transfer the signing ability from human user to AI proxy; (iii) There is a lack of an authentication method that supports tracing both human-driven and AI-driven avatars.

For the first challenge, current authentication methods mainly focus on establishing identity models for human-driven avatars to ensure that the virtual avatar can be traced back to its physical manipulator\cite{Ryu2022MutMeta,Yang2023Trace,Zhang2023Anti}, while ignoring the identity model of AI-driven avatars. Ryu \textit{et al}. \cite{Ryu2022MutMeta} built a three-factors identity based on account password, fingerprint feature, and public key to achieve login authentication between human-driven avatar and platform server. Yang \textit{et al}. \cite{Yang2023Trace} established a two-factors identity model based on user's iris feature and public key to guarantee the traceability of human-driven avatars. Zhang \textit{et al}. \cite{Zhang2023Anti} introduced the first impression of friends as an auxiliary factor to defend against the disguise attacks by human-driven avatars. The above identity models are only suitable for human-driven avatars and cannot guarantee the verifiability and traceability of AI-driven avatars. Therefore, it is necessary to construct an identity model to ensure that an AI-driven avatar is associated with its original manipulator.

For the second challenge, traditional authentication methods design signature algorithms to verify human-driven avatars in different scenarios\cite{George2019Third,Li2023Vibhead,Huang2023MutAuth,Thakur2023ECC}, while ignoring the need to transfer signing ability from human users to AI proxies. Thakur \textit{et al}. \cite{Thakur2023ECC} designed a signature algorithm based on elliptic curve cryptography(ECC) to verify human-driven avatars when human users logs into a metaverse platform. In Yang's framework \cite{Yang2023Trace}, they designed a chameleon collision signature scheme to verify the virtual and physical identities of human-driven avatars. But for AI-driven avatars, this authentication method have not yet focused on how to transfer the signing ability of human users. Therefore, it is necessary to build a transferable signature scheme to ensure that the signing ability of human users can be transferred to AI proxies.

For the third challenge, traditional authentication method \cite{Yang2023Trace} mainly focuses on the problem of tracing a human-driven avatar back to corresponding physical manipulator, while ignoring the traceability of AI-driven avatars. Aiming at this flaw, attackers may delegate someone's avatar to an AI proxy and carry out impersonation attacks without worrying about being tracked. Therefore, there is an urgent need to design a traceable authentication method that traces an avatar back to its original manipulator whether the avatar is driven by a human user or an AI proxy. 

\subsection{Main Challenge}

To solve the above problems, the main challenge is to design a proxy signature scheme that meets the three metrics of security, efficiency, and short signature length. Even though traditional schemes related to proxy signature pay extensive attention to various metrics  \cite{Hwang2003Threshold,Boldyreva2012Secure,Lin2018Short,Verma2019CBPS,Verma2020CBPS,Wang2022Params,Qiao2022ProxTII}, they fail to meet the above three metrics at the same time.

In response to the lack of rigorous security analysis, Boldyreva \textit{et al}. \cite{Boldyreva2012Secure} analyzed the security of some well-known proxy signature schemes and made appropriate modifications to make them provably secure. To address the signature delay in the UAV and IoT, Verma \textit{et al}.  \cite{Verma2019CBPS,Verma2020CBPS} designed two proxy signature schemes with low computational costs based on the ideas of BLS short signature \cite{Boneh2004BLS}. Regarding the problem of private verification \cite{Verma2019CBPS}  and the reuse attacks of original signature \cite{Verma2020CBPS}, Qiao \textit{et al}. \cite{Qiao2022ProxTII} constructed three proxy signature schemes to guarantee the public verifiability and unforgeability while ensuring efficiency.

Though the scheme \cite{Qiao2022ProxTII} meets the requirements of security and efficiency, its proxy signatures involve several parameters, resulting in longer signature length. If this scheme is applied in interactive environments such as large-scale avatars, it will bring a huge communication overhead to the authentication system. Therefore, how to build a proxy signature scheme that meets the three metrics is a major challenge facing by the traceable authentication method.

\subsection{Our Contributions}

Inspired by the short signature \cite{Boneh2004BLS} and the chameleon collision signature \cite{Yang2023Trace}, we propose a chameleon proxy signature scheme meeting the three metrics to transfer the signing ability of human users to AI proxies. Based on the proposed signature scheme, we design three authentication protocols for avatars to achieve login authentication, delegate authentication, and mutual authentication. These protocols  guarantee the virtual-to-physical traceability and defend against false accusation. To sum up,  the main contributions are as follows:

\begin{itemize}
     \item We construct a user's identity model that is compatible with both human-driven and AI-driven avatars. This model combines the iris feature of human manipulator and the metaverse public key of AI proxy to ensure that an AI-driven avatar is associated with its original manipulator.

     \item We propose a chameleon proxy signature scheme that meets the three metrics of security, efficiency and short signature length. The scheme is inspired the BLS short signature \cite{Boneh2004BLS} and the chameleon collision signature\cite{Yang2023Trace} to generate short proxy signatures in an efficient way.
     
    \item We designed three authentication protocols that can be used for login authentication, delegation authentication, and mutual authentication to provide conditions for virtual-to-physical tracing.

    \item We implement a simulated authentication system for avatars. The system not only completes mutual authentication between human-driven and AI-driven avatars,  but also achieves virtual-to-physical tracing on these two type of avatars. All the above operations are completed in about 1s, meeting the needs of metaverse applications.     
\end{itemize}

The rest of the paper is organized as follows. We introduce the background in the next section. The avatars authentication framework is presented in Section \ref{sec:AuthFramework}. Section \ref{sec:CPS} proposes the chameleon proxy signature. Section \ref{sec:AuthProto} designs the authentication protocols. The performance evaluations are given in Section \ref{sec:ImpEval}. We discuss this work and the future work in Section \ref{sec:Discu}.  Finally, we conclude the paper in Section \ref{sec:Conclu}.

\section{Background}
In this section, we present the relevant background, including avatars in metaverse, tracing human-driven avatars, and chameleon collision signature, to provide references for constructing avatars authentication framework.

\subsection{Avatars in Metaverse} \label{sec:Avatars}
\begin{figure}[htbp]
    \centering
     \subfigure[The human-driven avatar]{
     		\label{fig:Human_Avatar}    \includegraphics[width=0.40\textwidth]{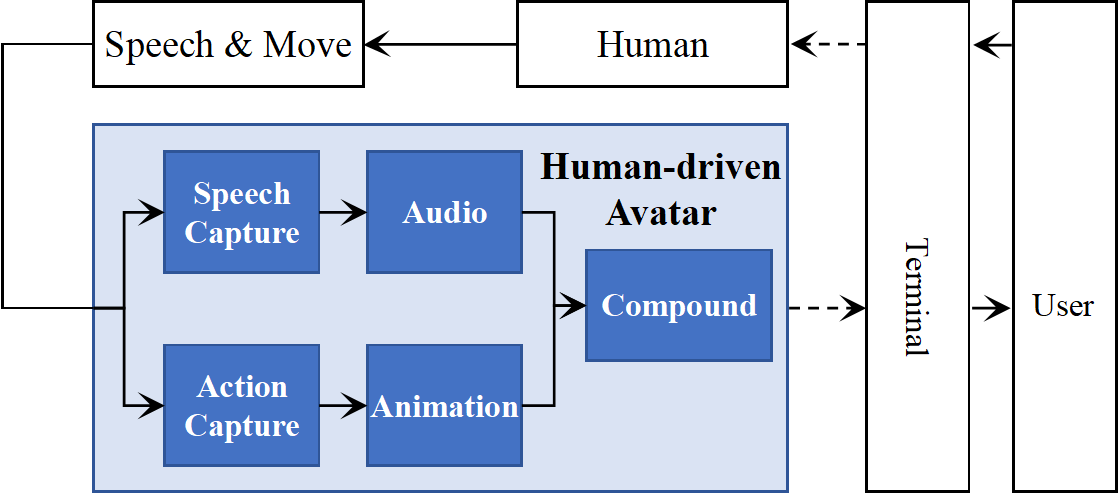}    }    \quad
    \subfigure[The AI-driven avatar]{
    		\label{fig:AI_Avatar}    \includegraphics[width=0.40\textwidth]{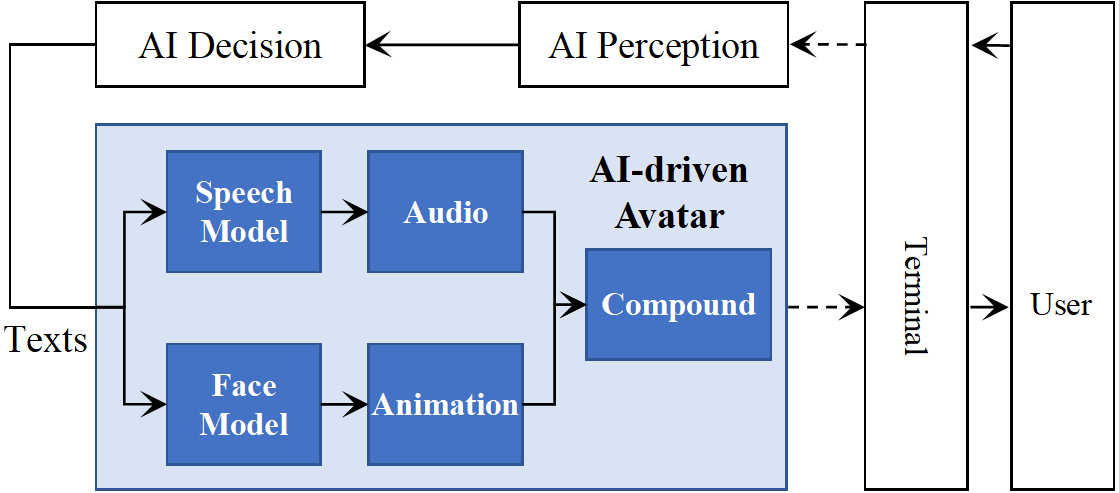}    }    \quad
    \caption{The generation process for different types of avatars.}
    \label{fig:Avatars}
\end{figure}

There are two types of avatars in the metaverse: human-driven avatars and AI-driven avatars. Though different metaverse platforms utilize different technologies to generate avatars \cite{Wirth2021Gait,Gu2024RealAva,Zhang2023RealAva,Basyoni2023AIGC}, their technical frameworks can be summarized as Fig.\ref{fig:Avatars}.

\textbf{Human-driven avatar} is the visual appearance of a physical manipulator in metaverse. The generation process of a human-driven avatar is shown as Fig.\ref{fig:Human_Avatar}. During a human manipulator speaking and moving, the wearable devices attached to the manipulator capture the user's voices and actions to form audio and animation \cite{Wirth2021Gait}, which are rendered as a visual avatar by the terminal devices of an interaction partner\cite{Gu2024RealAva}. The avatar's voices and actions are synchronized with the manipulator, providing an immersive experience for the partner \cite{Zhang2023RealAva}. In this light, a human-driven avatar is the unity of virtual identity and physical identity, where the virtual identity is the visual appearance in metaverse and the physical identity is the physical manipulator in the real world. 

\textbf{AI-driven avatar} is the visual appearance of artificial intelligence generated content (AIGC) \cite{Basyoni2023AIGC}, which is driven by an AI proxy program. A vivid example of AI-driven avatar in metaverse is shown as an AI shopkeeper\footnote{\href{https://www.youtube.com/watch?v=iZ_20vK94hc}{https://www.youtube.com/watch?v=iZ\_20vK94hc}} released by NVIDIA, which freely converses with players and provides game guidance. The generation process of AI-driven avatars can be explained by the Fig.\ref{fig:AI_Avatar}. When the AI proxy program receives information from an external terminal, the AI proxy calls the intelligent perception and intelligent decision-making modules to generate response texts, and then calls the avatar engine to generate corresponding audio and animation based on a user's AI models\footnote{\href{https://build.nvidia.com/nvidia/audio2face}{https://build.nvidia.com/nvidia/audio2face}}, which are rendered into a visual avatar for other users.

\subsection{Tracing Human-driven Avatars}
The consistency between avatar's virtual and physical identities is the key to guaranteeing the virtual-to-physical traceability. Yang \textit{et al}. \cite{Yang2023Trace} proposed a traceable authentication framework for human-driven avatars. Before entering the metaverse, users provide his/her actual identity $ID$, public key $pk$ and biometric template $T$ to IdP, who publishes corresponding metaverse identity token $MIT=\{SN,M,pk,T\}$ for users. Among them, the serial number $SN$ and user's $ID$ are stored in a trusted database to support virtual-to-physical tracing. During the authentication process between avatars, an avatar acting as a verifier periodically obtains and verifies the virtual identity $VID$ and the physical identity $PID$  of another avatar acting as a prover based on $MIT$, ensuring that the prover's virtual identity is consistent with its physical identity. If the prover acts maliciously, the verifier submits these identity parameters $\{MIT, VID,PID\}$ to IdP, who checks these parameters and tracks down the prover's manipulator based on $(SN, ID)$.

\subsection{Chameleon Collision Signature}
Chameleon signature is a changeable signature algorithm, guaranteeing the consistency between the avatar's virtual identity and physical identity. Since a chameleon collision is forged by the user's private key, a new collision can be treated as a chameleon signature related to the old collision. Based on this idea, Yang \textit{et al}. \cite{Yang2023Trace} propose an efficient chameleon collision signature to reduce the key holding cost and signing overhead. This signature scheme consists of the following parts:

\begin{itemize}

\item $Setup(\mathcal{K})\rightarrow Parms$. The input of this probabilistic algorithm is a security parameter $\mathcal{K}$ and the output is the system parameter $Parms$.

\item $KeyGen(Parms)\rightarrow (pk,sk) $. The key generation algorithm takes the system parameter $Parms$ as the input and outputs the public-private key pair $(pk, sk)$.

\item $Hash(pk,M) \rightarrow (h,R)$. The hash algorithm takes $pk$ and a message $M$ as input and outputs the chameleon hash value $h$ and the check parameter $R$ of $M$.

\item $Check(pk,h,M,R)\rightarrow b\in\{0,1\}$. The compatibility detection algorithm takes as input the chameleon triplet$(h,M,R)$ and $pk$. It outputs a decision $b\in\{0,1\}$ indicating whether the $( pk, h, M, R )$ is compatible or not. 

\item $Sign(sk,h,M,R,M^\prime)\rightarrow R^\prime$. To sign a message $M^\prime$, the algorithm takes as input $sk$ , $(h,M,R)$, and $M^\prime$. It outputs the check parameter $R^\prime$ of $M^\prime$, where the pairs $(M,R)$ and $(M^\prime, R^\prime)$ are called a colliding signature with respect to the hash value $h$.

\item $Verify(pk,h,M,R,M^\prime,R^\prime) \rightarrow b $.  To verify the colliding signatures $(M,R)$ and $(M^\prime, R^\prime)$ with respect to $h$, the algorithm detects the compatibility of $(h,M,R)$ and $(h,M^\prime,R^\prime)$ using the $Check$ algorithm. It outputs a decision $b\in\{0,1\}$ indicating whether the pairs $(M,R)$ and $(M^\prime, R^\prime)$ form a valid signature or not.

\end{itemize}

\section{Avatars Authentication Framework}\label{sec:AuthFramework}
\begin{figure}[htbp]
\begin{center}
    \includegraphics[width=0.48\textwidth]{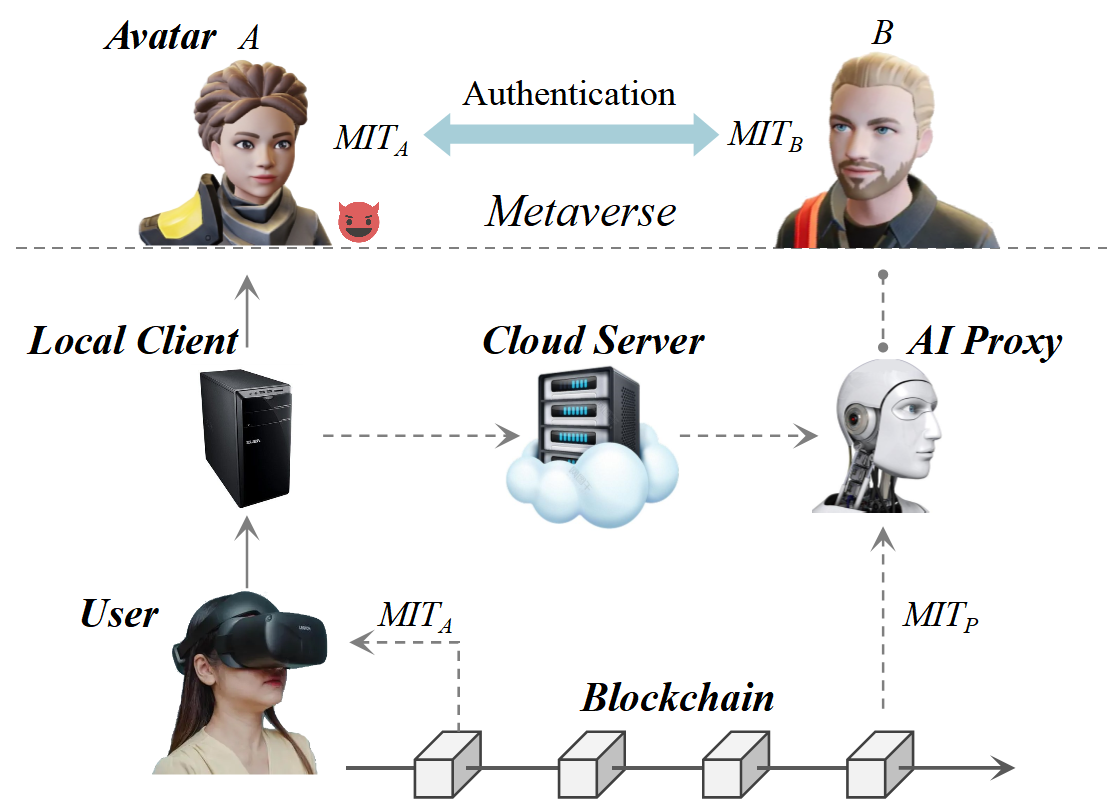}
    \caption{\small{The traceable authentication framework.}}
    \label{fig:Framework}
 \end{center}
\end{figure}

The virtual-to-physical traceability is the key to mitigating the impersonation attacks by AI-driven avatars. In this section, we construct an authentication framework as shown in Fig.\ref{fig:Framework} to guarantee the traceability including both the human-driven and AI-driven avatars.

\subsection{System Model}

\begin{itemize}

\item $\boldsymbol{Avatar}$ is the virtual appearance of a driven entity, which can be a human user or an AI proxy program. To ensure the uniqueness of physical identity, we require that an avatar can only be driven by one entity in the same time.

\item $\boldsymbol{User}$ is the physical manipulator of a human-driven avatar and is also the original manipulator of an AI-driven avatar. Before entering the metaverse, users need to register in an IdP to obtain corresponding MIT, by which users are able to create arbitrary avatars. 

\item  $\boldsymbol{Local \; Client}$ is the main carrier of metaverse applications, which can be a personal computer (PC),  a smartphone (SP), or a low-computation-power device (LCPD). The client communicates with cloud server and user's personal devices, such as head mount display (HMD), to process various interactive information.

\item $\boldsymbol{Cloud \; Server}$ deploys a metaverse platform and AI engines, such as the avatar cloud engine of NVIDIA. The server provides conditions for users to transfer their signing ability to AI proxies.

\item  $\boldsymbol{AI\; Proxy}$ is an proxy program with specific interaction tasks. It accepts users' delegation and calls AI engines to generate visual avatars, providing immersive interaction for other users. During the delegation process, users need to provide AI proxy with their AI models and physical features, such as iris, to guarantee the traceability of AI-driven avatars.

\item $\boldsymbol{Blockchain}$ is a decentralized storage system independent of cloud servers. It is only responsible for storing public parameters related to users' identities such as $MIT$ to ensure the persistence of user data.

\end{itemize}

\subsection{User's Identity Model}
In this part, we construct a user's identity model based on human user's iris feature and AI proxy's public key, ensuring an AI-driven avatar is associated with its original manipulator.

\begin{figure}[htbp]
\begin{center}
    \includegraphics[width=0.40\textwidth]{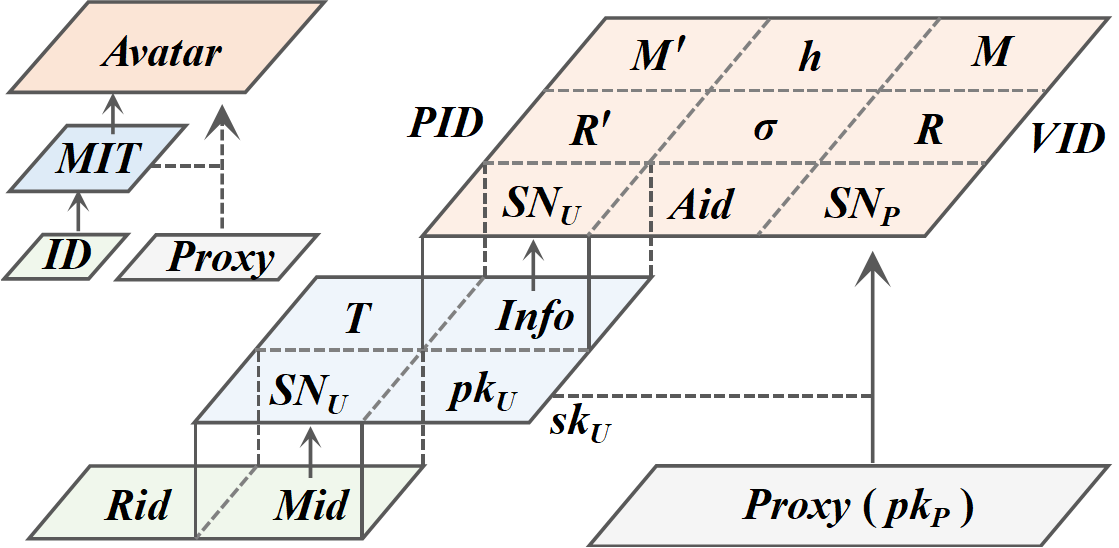}
    \caption{\small{The user's identity model.}}
    \label{fig:IdenMod}
 \end{center}
\end{figure}

The user's identity model is a security enhancement measure over the avatar's identity model \cite{Yang2023Trace}. The constructed model  $User=\{ID,MIT,Avatar\}$  is shown as the Fig.\ref{fig:IdenMod}, in which $ID=(Rid,Mid)$ is the user's identification number related to the real world and the  metaverse, $MIT=( SN_U, pk_U, T, Info)$ is the metaverse identity token connecting the physical manipulator and the digital avatar, and $Avatar=( SN_U,Aid, SN_P, \sigma, h, VID, PID)$ is the digital appearance in metaverse. All parameters and the corresponding meanings are shown in the TABLE \ref{tab:Notatins}. 

In this model, $SN_U$ and $SN_P$ imply the avatar's driver type, meaning that the avatar is driven by a physical manipulator when $SN_U==SN_P$; otherwise, the avatar is driven by an AI proxy. $VID=(M,R)$ is the avatar's virtual identity, where $M$ is the detailed description of the avatar including face model and speech model, $R$ is the check parameter of $M$.  $PID=(M^\prime,R^\prime)$ is the avatar's physical identity, where $M^\prime$ is the iris feature taken from the avatar's manipulator, $R^\prime$ is the check parameter of $M^\prime$.

During the delegation process, the manipulator generates an original signature on his/her iris feature and the AI proxy's public key to form the avatar's physical identity, where the proxy's public key indicates that the manipulator transfers his/her signing ability to the AI proxy. After the proxy verifies the physical identity, it generates a corresponding proxy signature about $M$ to form the avatar's virtual identity, indicating that the AI proxy has the signing ability of the avatar.

\begin{table}[htbp]
 \small
  \begin{center}

  \caption{User's identity parameters and meanings}
  \label{tab:Notatins}
 \begin{tabular}{p{50 pt}<{\centering}p{180 pt}<{\centering}}
        \hline
        \specialrule{0em}{1pt}{1pt}%改变行间距
        Symbol & Description  \\
        \specialrule{0em}{1pt}{1pt}%改变行间距
        \hline     
        \specialrule{0em}{1pt}{1pt}%改变行间距     
        $ID$ & The user's identification number \\
        $Rid$ & The unique identity in the real world  \\ 
        $Mid$ & The unique identity in the metaverse \\ 
        $MIT$ & The metaverse identity token\\ 
        $SN_U$ & The serial number in user's $MIT$\\
        $pk_U$ & The user's public key   \\    
        $T$ & The iris template of user\\
        $Info$ & The anonymous description of the user   \\
        $Avatar$ & The digital appearance in metaverse\\ 
        $Aid$    &  The avatar's identification number\\
        $SN_P$ &  The serial number in AI proxy's $MIT$\\
        $\sigma$    &   The signature \\
        $h$ &  The chameleon hash\\
        $VID$  &   The visual virtual identity\\
        $PID$ &  The traceable physical identity\\        
        $Proxy$  &   The AI proxy\\
         \specialrule{0em}{1pt}{1pt}%改变行间距
        \hline
  \end{tabular}
  \end{center}
\end{table}

\section{Chameleon Proxy Signature}\label{sec:CPS}
Traditional proxy signature schemes \cite{Verma2019CBPS,Qiao2022ProxTII} fail to meet the three metrics of security, efficiency, and short signature length, bringing a huge computational cost and communication overhead to the authentication system. In this section, we design a chameleon proxy signature scheme to generate short proxy signatures in an efficient way.

\subsection{Instantiation of CPS}\label{sec:ChamSign}
The BLS short signature \cite{Boneh2004BLS} generates signatures with  shorter length. The chameleon collision signature \cite{Yang2023Trace} ensures the consistency between avatar's virtual and physical identities. Making full use of these advantages,  we propose a chameleon proxy  signature scheme  $CPS = \{Setup$, $KeyGen$, $Hash$, $DGen$, $PVer$, $PSig\}$ as shown in the following:

\begin{itemize}

\item $Setup(\mathcal{K})\rightarrow Parms$. Let $\mathcal{K}$ be a security parameter in the signature system. The notations $\mathbb{G}$ and $\mathbb{G}_T$ are multiplicative cyclic groups of prime order $q\geq 2^\mathcal{K} $, where $g$ is a generator of $\mathbb{G}$. The pairing  $e:\mathbb{G}\times\mathbb{G}\rightarrow \mathbb{G}_T$ is an eﬃciently computable bilinear map and $\mathbb{Z}_q$ is a finite field of order $q$, where the pairing satisfies $e(g^a,g^b)=e(g,g)^{ab}$  for any $a,b\in \mathbb{Z}_q$. The system selects a global anti-collision hash functions $H:\{0,1\}^*\rightarrow\mathbb{G}$ mapping bit strings of arbitrary length to an elements in $\mathbb{G}$. Finally, the algorithm publishes the system parameters as
$$Parms=\{\mathbb{G},\mathbb{G}_T,g_,q,e,H\}.$$
 
\item $KeyGen(Parms)\rightarrow (pk,sk) $.  The key generation algorithm takes $Parms$ as input. The algorithm selects a randomness $x\stackrel{R}{\leftarrow}  \mathbb{Z}_q$ as the private key $sk$ and calculates $y=g^x \in \mathbb{G}$ as the public key $pk$, where the symbol ``$\stackrel{R}{\leftarrow}$'' means to randomly select an element from a set. It outputs the public-private key pair as	
$$sk = x, \; pk = y. $$
\item $Hash(M,pk) \rightarrow (h,R)$. The algorithm takes as input the public key $pk=y$ and a message $M\in \{0,1\}^*$. It outputs the chameleon hash value $h$ and the corresponding check parameter $R$ of $M$ as
\begin{equation}\nonumber \label{equ:Hash}
\begin{split}
     m=H(M),\; r\stackrel{R}{\leftarrow} \mathbb{Z}_q,\\
     h=m \cdot y^r,  \; 	R=g^r.
\end{split}
\end{equation}

\item $DGen(sk_A, M, pk_B) \rightarrow (\sigma_A, h_B, R) $. To generate an original signature about message $M$, the algorithm takes as input the private key $ sk_A = x_A $ and the public key $pk_B=y_B$. It outputs the tuple $ (\sigma_A, h_B, R) $ as 
\begin{gather*}
(h_B, R) = Hash(M,pk_B),\\
\sigma_A=h_B^{x_A}. 
\end{gather*}
Among them, $M$ and $ (\sigma_A, h_B, R) $ form the original signature $(\sigma_A, h_B, M, R)$.

\item $PVer(pk_A, \sigma_A, h_B, M, R, pk_B) \rightarrow b $. To check a signature $(\sigma_A, h_B, M, R)$, the algorithm takes as input the public keys  $pk_A=y_A$ and $pk_B=y_B$. It checks the compatibility as
\begin{gather*}   
    e(\sigma_A,g) \stackrel{?}{=} e(h_B,y_A), \tag{3.1} \label{eq:PV1} \\ 
    e(h_B/m,g) \stackrel{?}{=} e(R, y_B). \tag{3.2} \label{eq:PV2}
\end{gather*}
Among them, the symbol ``$\stackrel{?}{=}$'' indicates whether an equation is hold or not. If  the equations \eqref{eq:PV1} and \eqref{eq:PV2} are both hold, the algorithm outputs $b=1$; otherwise it outputs $b=0$.

\item $PSig(sk_B,h_B, M^\prime) \rightarrow R^\prime $.  To generate a proxy signature about the message $M^\prime$ under the chameleon hash $h_B$, the algorithm inputs $sk_B=x_B$ and outputs the check parameter $R^\prime$ as
\begin{gather*}   
    m^\prime = H(M^\prime), \; R^\prime=(h/m^\prime)^{(1/x_B)}. 
\end{gather*}
Among them, $R^\prime$ and  $ (\sigma_A, h_B, M^\prime) $ form the proxy signature $(\sigma_A$, $h_B$, $M^\prime$, $R^\prime)$.

\end{itemize}

\subsection{Security of CPS}
For the proposed signature scheme CPS, an attacker may forge the signatures related to an original signer or a proxy signer. The security of CPS depends on the CDH assumption and the DCDH assumption\cite{Bao2003Variations}. In the following, we first define the unforgeability of the original signature under the adaptive chosen message attack (OS-EUF-CMA), and then define the unforgeability of the proxy signature under the adaptive chosen message attack (PS-EUF-CMA).  Details of the proof are given in Appendix \ref{sec:SecuProof}.

\textbf{Definition 1 (The Unforgeability of Original Signature).} We say that the chameleon proxy signature satisfies original signature unforgeability under adaptive chosen message attacks (OS-EUF-CMA), if there is no polytime adversary who wins the game $Exp_{CPS,\mathcal{A}}^{OS\text{-}EUF}(\mathcal{K})$ with a non-negligible advantage.

\textbf{Theorem 1}. If the CDH assumption holds on $\mathbb{G}$, the chameleon proxy signature is OS-EUF-CMA.

\textbf{Definition 2 (The Unforgeability of Proxy Signature).} We say that the proposed signature scheme satisfies proxy signature unforgeability under adaptive chosen message attacks (PS-EUF-CMA), if there is no polytime adversary who wins the game $Exp_{CPS,\mathcal{B}}^{PS\text{-}EUF}(\mathcal{K})$ with a non-negligible advantage.

\textbf{Theorem 2}. If the DCDH assumption holds on $\mathbb{G}$, the chameleon proxy signature is PS-EUF-CMA.

\section{Authentication Protocols}\label{sec:AuthProto}
In this section, we design three authentication protocols for avatars based on the proposed signature scheme to guarantee the virtual-to-physical traceability and defend against the false accusation.

\subsection{Construction of Protocols}
In metaverse,  users are able to delegate their avatars to AI proxies, which generate the corresponding AI-driven avatar on behalf of the original manipulator to interact with other users. Based on this feature, firstly, we design a login authentication protocol to implement human-driven avatars login. Secondly, we design a delegation authentication protocol to transfer the signing ability of human user to AI proxy. Finally, we design a mutual authentication protocol for human-driven and AI-driven avatars to verify each other.

\begin{figure*}[htbp]
    \centering
     \subfigure[Login authentication protocol]{
     		\label{fig:Pro_1}    \includegraphics[width=0.22\textwidth]{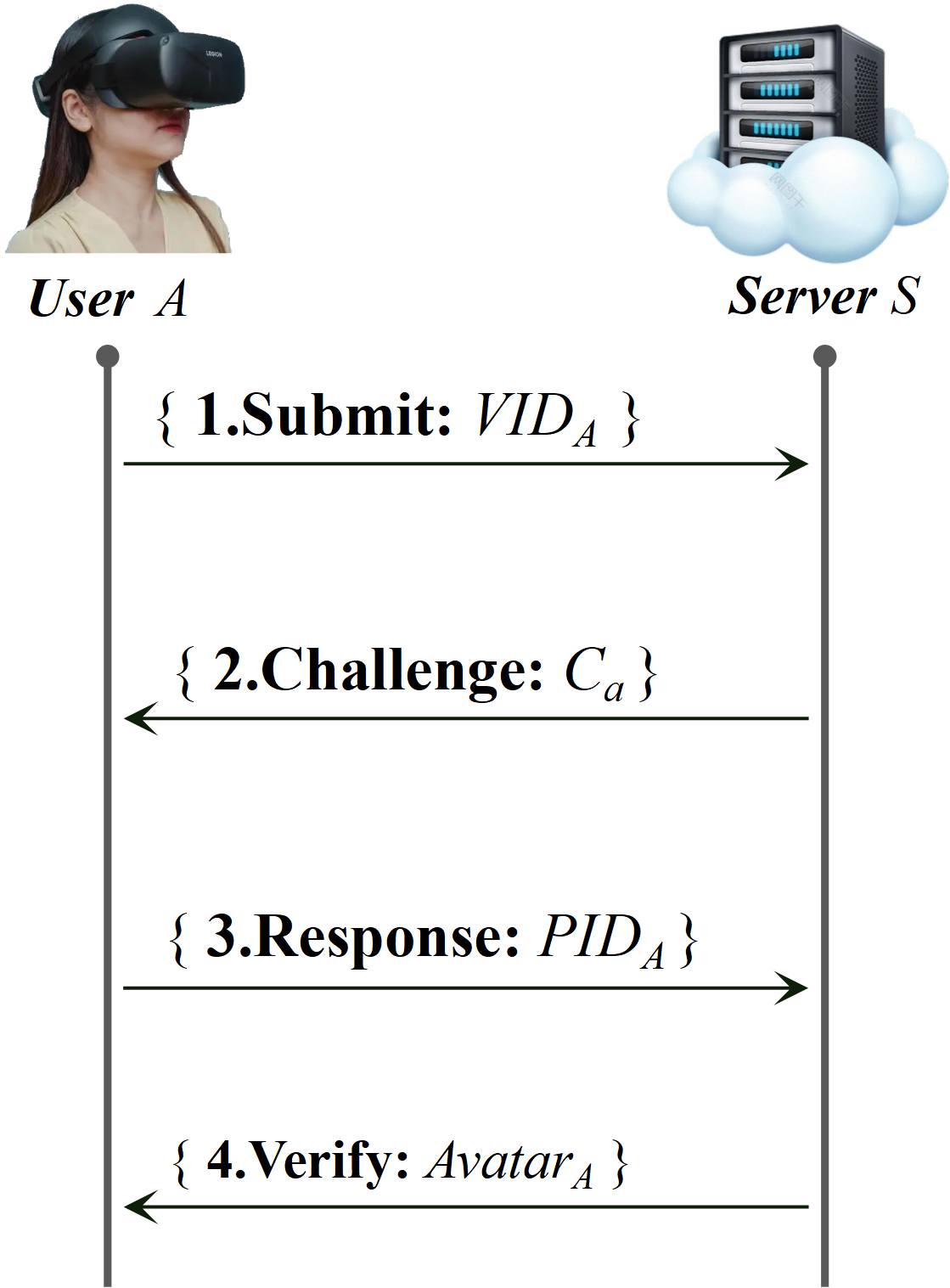}    }    \quad
    \subfigure[Mutual authentication protocol]{
    		\label{fig:Pro_2}    \includegraphics[width=0.39\textwidth]{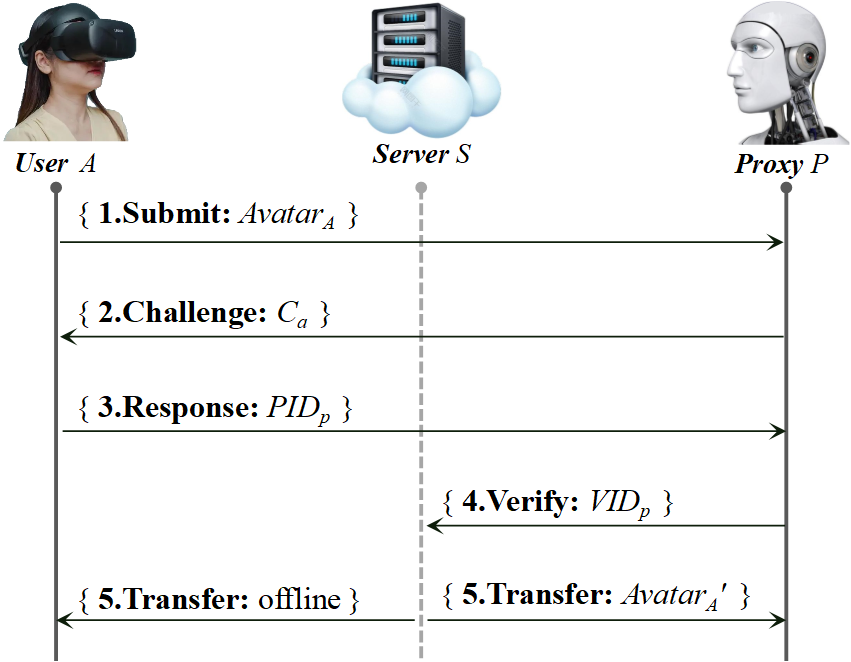}    }    \quad
    \subfigure[Delegation authentication protocol]{
    		\label{fig:Pro_3}    \includegraphics[width=0.23\textwidth]{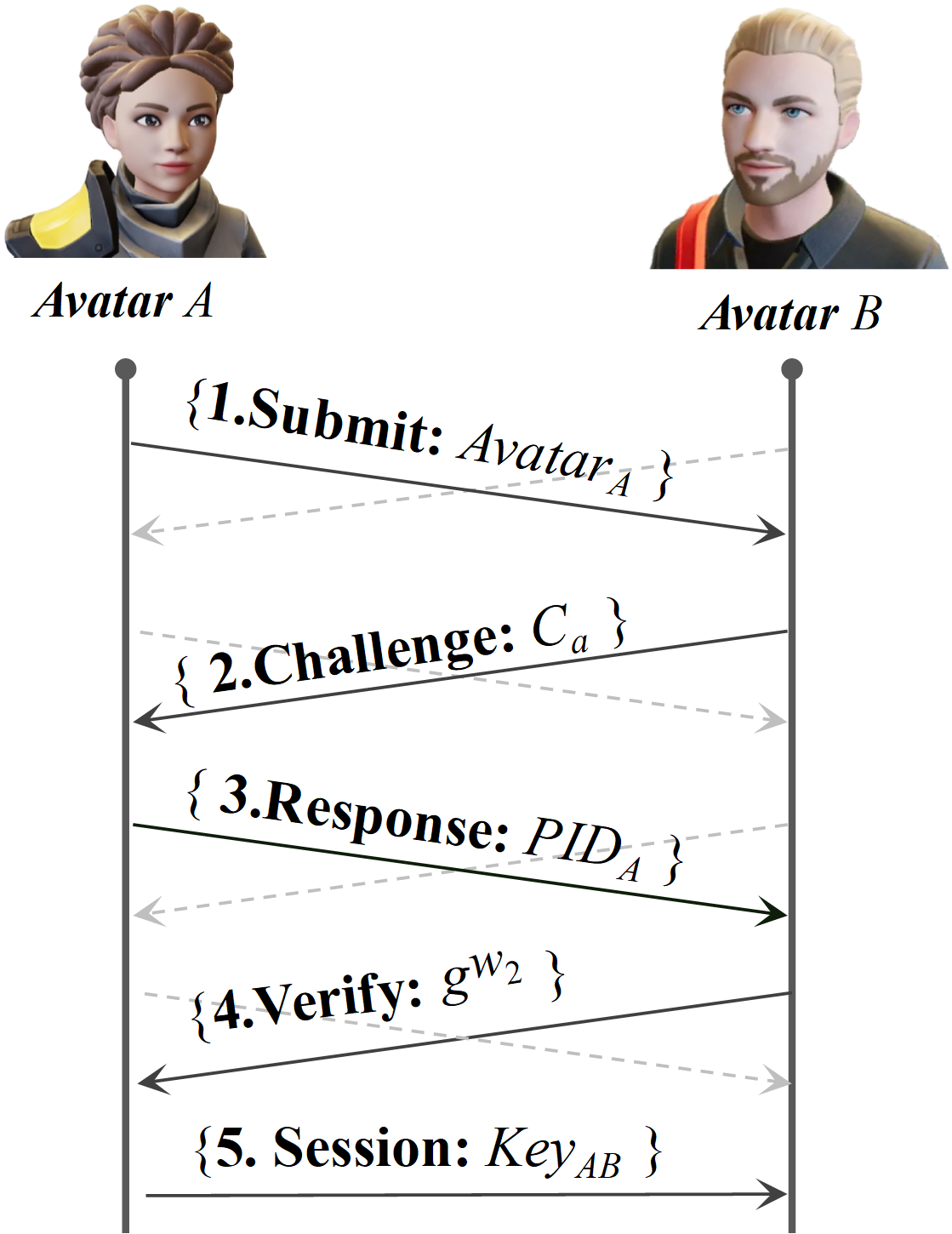}    }    \quad
    \caption{The authentication protocols for avatars.}
\end{figure*}

\subsubsection{Login Authentication Protocol}
The login authentication protocol assumes that the user $A$ has registered an avatar in the cloud server $S$. The protocol consists of four phases as shown in Fig.\ref{fig:Pro_1}: (1) $A$ submits the avatar's virtual identity $VID_ A$ to $S$ to initiate a login claim; (2) $S$ throws a random challenge $C_a$ to the user after verification the $VID_A$; (3) $A$ samples an iris feature to construct the avatar's physical identity $PID_ A$ as a response; (4) $S$ creates the corresponding $Avatar_A$ for the user $A$ after verification the $PID_A$.

During the login process, $A$ generates the original signature $(\sigma_A,h_A,R_A) \leftarrow DGen(sk_A,M_A,pk_A)$ to construct $VID_A=(M_A,R_A)$, where $pk_A$ means that $A$ delegates his/her  signing ability to himself/herself. To construct $PID_A$, $A$ generates the proxy signature $R_A^\prime \leftarrow PSig(sk_A,h_A, M_A^\prime)$ about the iris feature $M_A^\prime$, where $h_A$ ensures that the avatar's $PID_A$ is consistent with its $VID_A$. The details of login authentication protocol are given in Appendix \ref{sec:LoginProto}.

\subsubsection{Delegated Authentication Protocol}
The delegated authentication protocol allows a human user $A$ to transfer his/her signing ability to an AI proxy $P$. The protocol consists of five phases as shown in Fig.\ref{fig:Pro_2}: (1) $A$ provides $P$ with the full information of the avatar $Avatar_A$ to initiate a delegation; (2) $P$ verifies $Avatar_A$'s virtual identity and throws a random challenge to $A$; (3) $A$ generates the physical identity $PID_P$ about the proxy's public key $pk_P$ as a response; (4) $P$ verifies the $PID_P$ and generates a corresponding virtual identity $VID_P$ for $S$; (5) $S$ updates $Avatar_A$ to $Avatar_A^\prime$ for $P$ and forces $A$ offline.

During the delegation process, $A$ generates the original signature $(\sigma_A^\prime,h_P,R_P^\prime) \leftarrow DGen(sk_A,M_A^\prime, pk_P)$ about his/her iris feature $M_A^\prime$ to construct $PID_P=(M_A^\prime,R_A^\prime)$, while $P$ generates a proxy signature $R_P \leftarrow PSig(sk_P, h_P, M_A)$ to construct the $VID_P=(M_P,R_P)=(M_A,R_P)$. Among them, the $PID_P$ ensures that the physical identity of the AI-driven avatar is associated with its original manipulator. The details of delegated authentication protocol are given in Appendix \ref{sec:DelegProto}.

\subsubsection{Mutual Authentication Protocol}
The mutual authentication protocol achieves verification between human-driven and AI-driven avatars to establish a session key, in which both parties verify each other's virtual and physical identities in the same way as shows in Fig.\ref{fig:Pro_3}. Here, we let the avatar $A$ as a prover and let the avatar $B$ as a verifier to introduce the protocol: (1) $A$ submits the full information of $Avatar_A$ to $B$; (2) $B$ verifies the $VID_A$ based on $A$'s driver type and throws a random challenge; (3) $A$ signs iris feature to form a physical identity $PID_A$ as a corresponding response; (4) $B$ verifies $PID_A$ and sends a random parameter to $A$; (5) $A$ establishes a session key based on the random parameter.

It is worth noting that $B$ determines $A$'s driver type based on $SN_U \stackrel{?}{=}$ $SN_P$ and selects corresponding public keys to verify $A$, achieving mutual authentication between human-driven and AI-driven avatars. Moreover, $B$ retains $A$'s identity parameters to support virtual-to-physical tracing. The details of mutual authentication protocol are given in Appendix \ref{sec:MutualProto}.
 
\subsection{Security of Protocols}
The security goal of the proposed authentication protocols is to guarantee the virtual-to-physical traceability and defend against false accusation. The key to guaranteeing traceability is that both the human-driven and AI-driven avatars must submit the iris feature of original manipulator to verifier. The key to defending against false accusation is that the identity parameters generated by provers are unforgeability. In the following, we analyze the protocols' security including virtual-to-physical traceability and defending against false accusation.

\subsubsection{Virtual-to-Physical Traceability}
The traceability considered in this work is divided into two categories: tracing a human-driven avatar back to its physical manipulator and tracing an AI-driven avatar back to its original manipulator. During the tracing process, the reporter submits visual interaction images, such as screenshots, and the full identity parameters of an avatar to the IDP, who verifies these parameters and tracks down the manipulator. Among them, the interaction images are associated with the avatar's virtual identity. 

The traceability of human-driven avatars lies in the mutual authentication protocol. In the process of mutual authentication  between a human-driven avatar $Avatar_A$ and an avatar $B$, $B$ verifies the $VID_A$ and $PID_A$ of $Avatar_A$ to obtain the iris feature of manipulator, which forms the interactive evidence related to the physical manipulator. If the manipulator behaves maliciously, $B$ takes a screenshot and submits all evidences to the IdP, who verifies these parameters and traces the $Avatar_A$ back to its physical manipulator based on $(SN_A,ID_A)$, where $SN_A$ implies in $Avatar_A$. Therefore, the mutual authentication protocol guarantees the virtual-to-physical traceability of human-driven avatars.

The traceability of AI-driven avatars lies in the delegated authentication protocol. During the delegation process between the original manipulator $A$ and an AI proxy $P$, $P$ obtains the physical identity $PID_P= (M_P^\prime,R_P^\prime)$ from $A$ to guarantee that the physical identity of the AI-driven avatar $Avatar_A^\prime$ associated with its original manipulator, where the $M_P^\prime$ is constructed by the iris feature of the original manipulator. In the process of mutual authentication, the $Avatar_A^\prime$ providers interaction partner with its $VID_P$ and  $PID_P$, by which the partner can trace the $Avatar_A^\prime$ back to its original manipulator with the help of IdP. Therefore, the delegated authentication protocol guarantees the virtual-to-physical traceability of AI-driven avatars.

In summary, the designed authentication protocols guarantee the virtual-to-physical traceability including both the human-driven and AI-driven avatars.

\subsubsection{Defending against False Accusation} 
For a malicious reporter, he may forge the identity parameters of a human-driven avatar or an AI-driven avatar to frame an honest user. 

$\cdot$ Case 1: For a human-driven avatar, the reporter has the full information of the legal avatar $Avatar_A=\{SN_A$, $Aid_A$, $SN_A$, $\sigma_A$, $h_A,VID_A,PID_A\}$ and a forged interaction image. The reporter's goal is to forge a proxy signature $(\sigma_A,h_A,M_A^*,R_A^*)$ such that $Verify(pk_A$, $\sigma_A$, $h_A$, $M_A^*$, $R_A^*,pk_A)=1$, where the avatar's $VID_A^*=(M_A^*$, $R_A^*)$ is associated with the forged image. According to the PS-EUF-CMA of the proposed signature scheme, the reporter fail to forge such parameters $(M_A^*,R_A^*)$. Therefore, the proposed protocol can defend against the false accusation on human-driven avatars.

$\cdot$ Case 2: For an AI-driven avatar, we allow the reporter to collude with an AI proxy to possess the proxy's private key. With this private key, the reporter can generate arbitrary $VID_P=(M_P,R_P)$ related to the forged image and arbitrary $PID_P=(M_P^\prime,R_P^\prime)$ related to the target manipulator's iris feature. Thus, the reporter's goal is to forge an original signature $(\sigma_A^*,h_P,M_P^\prime,R_P^\prime)$, such that $Verify(pk_A,\sigma_A^*,h_P,M_P^\prime,R_P^\prime
,pk_P)=1$. According to the OS-EUF-CMA of the proposed signature scheme, the reporter fail to forge such a parameter $\sigma_A^*$. Therefore, the proposed protocol can defend against the false accusation on AI-driven avatars.

In summary, the proposed authentication protocols can defend against the false accusations on human-driven and AI-driven avatars.

\section{Performance Evaluation}  \label{sec:ImpEval}

In this section, we design a simplified authentication system to evaluate the performance of the proposed signature scheme and the designed protocols. To simulate the interaction between human-driven and AI-driven avatars, we implement a simplified AI proxy program based on Java and install it on the three types of device platforms: PC, SP, and LCPD. The device parameters of different platforms are shown as the TABLE \ref{tab:Device}, where we simulate the device of LCPD with Raspberry Pi.

\begin{table}[htbp]
      \footnotesize
      \centering
      \caption{Parameters of different devices }
      \label{tab:Device}
      \begin{threeparttable}
      \begin{tabular}{p{50 pt}<{\centering}p{80 pt}<{\centering}p{35 pt}<{\centering}p{25 pt}<{\centering}}
          \hline
        \specialrule{0em}{1pt}{1pt}%改变行间距
        Type & Device & CPU & ARM  \\
        \specialrule{0em}{1pt}{1pt}%改变行间距
           \hline
          \specialrule{0em}{1pt}{1pt}%改变行间距
          HMD & HUAWEI VR Glass  & --  & --  \\
          PC & DELL Precision 3650 & 2.8 GHz  & 64 GB  \\
          SP & HUAWEI P40  & 2.8 GHz  & 8 GB  \\
          LCPD &  Raspberry Pi 4B & 1.5 GHz  & 4 GB \\
      \specialrule{0em}{1pt}{1pt}%改变行间距
        \hline
      \end{tabular}
\end{threeparttable}
\end{table}

\begin{table}[htbp]
\footnotesize
\centering
\caption{ The bit length of proxy signature schemes}
\label{tab:CommOver}
\begin{threeparttable}
    \begin{tabular}{p{40pt}<{\centering} p{80pt}<{\centering}  p{90pt}<{\centering}}
        \hline
         \specialrule{0em}{1pt}{1pt}%改变行间距
        Scheme & Original Signature & Proxy Signature \tnote{1} \\
        \hline
        \specialrule{0em}{1pt}{2pt}%改变行间距       
       Verma\cite{Verma2019CBPS} &	$ 1\,|G|  $  & 	$ 1 \, |G|$   \\
       \specialrule{0em}{1pt}{2pt}%改变行间距
       Verma\cite{Verma2020CBPS} &	$ 1\,|G|+1\,|Z|  $  & 	$ 2 \, |G|$   \\
       \specialrule{0em}{1pt}{2pt}%改变行间距
       Qiao\cite{Qiao2022ProxTII} &	$ 1\,|G| + 1\,|Z| $ &  $ 3 \, |G| + 1\,|Z|  $   \\
        \specialrule{0em}{1pt}{2pt}%改变行间距
        Yang\cite{Yang2020MBP} &	$ 4\,|G|+1\,|Z|  $  & 	$ 7 \, |G|+1\;|Z|$   \\               \specialrule{0em}{1pt}{2pt}%改变行间距
       Ours &	$ 3\,|G|  $  & $1\,|G|$  \\
        \hline
    \end{tabular}
      \begin{tablenotes}
        \footnotesize
        \item[1] We utilize $|G|$ and $|Z|$ to denote the bit length of elements in groups $\mathbb{G}$ and finite field $\mathbb{Z}_p$, respectively.
      \end{tablenotes}
\end{threeparttable}
\end{table}

\begin{table}[htbp]
\footnotesize
\centering
\caption{ The computation cost of the proxy signature}
\label{tab:ComputationCost}
\begin{threeparttable}
    \begin{tabular}{p{37pt}<{\centering} p{34pt}<{\centering} p{34pt}<{\centering} p{34pt}<{\centering} p{55pt}<{\centering}}
        \hline
         \specialrule{0em}{1pt}{1pt}%改变行间距
        Scheme & DGen & DVer & PSig & PVer \tnote{1} \\
        \hline
        \specialrule{0em}{1pt}{2pt}%改变行间距
       Verma\cite{Verma2019CBPS} &	$ 1\,E  $ &	$ 2 \, P  $  & $ 2 \, E  $ & 	$ 2 \, E + 1\,M+2\,P  $   \\
       \specialrule{0em}{1pt}{2pt}%改变行间距
       Qiao\cite{Qiao2022ProxTII} &	$ 1\,E + 1\,M $ &	$ 4 \, E + 3\,M  $  & $ 2\,E + 1\,M $ & 	$ 4 \, E + 3\,M  $   \\
        \specialrule{0em}{1pt}{2pt}%改变行间距
       Ours &	$ 3\,E + 1\,M $ &	$ 1\,M+4\,P $  &	 $ 2\,E + 1\,M $ & $1\,E + 1\,M + 4\,P$  \\
        \hline
    \end{tabular}
      \begin{tablenotes}
        \footnotesize
        \item[1] We utilize $E$, $M$, and $P$ to denote the exponential, multiplication, and bilinear map operation on $\mathbb{G}$, respectively.
      \end{tablenotes}
\end{threeparttable}

\end{table}

\begin{figure}[htbp]
\begin{center}
    \includegraphics[width=0.48\textwidth]{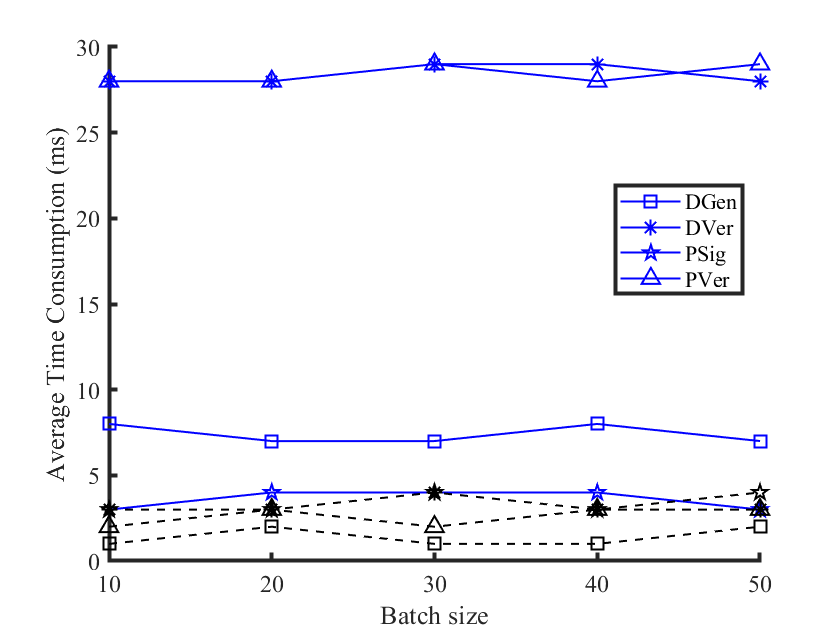}
    \caption{\small{The time consumption of CPS. The solid line and the dotted line in the figure are the time consumption of our scheme and that of Verma, respectively.}}
    \label{fig:ProxySign}
 \end{center}
\end{figure}

\begin{figure*}[htbp]
    \centering
    \subfigure[Login authentication protocol]{
     		\label{fig:LoginAuth}    \includegraphics[width=8.3cm]{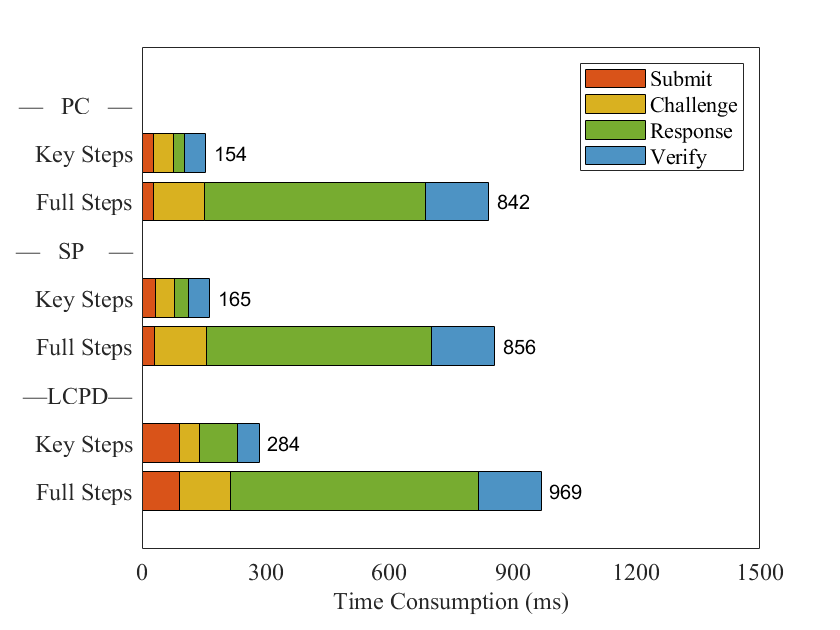}    }    \quad
     \subfigure[Delegation authentication protocol]{
     		\label{fig:DlegAuth}    \includegraphics[width=8.3cm]{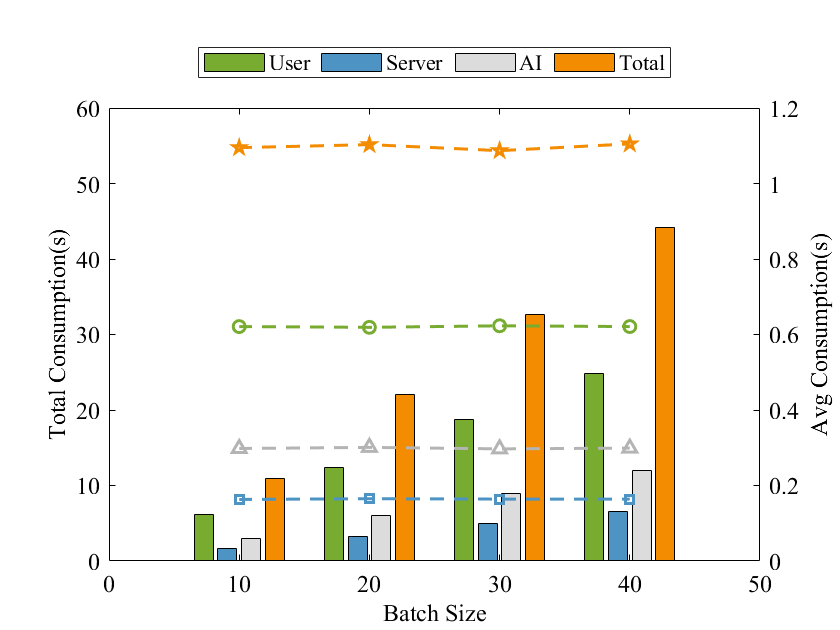}    }    \quad
    \subfigure[Mutual authentication protocol]{
    		\label{fig:MutAuthen}    \includegraphics[width=8.3cm]{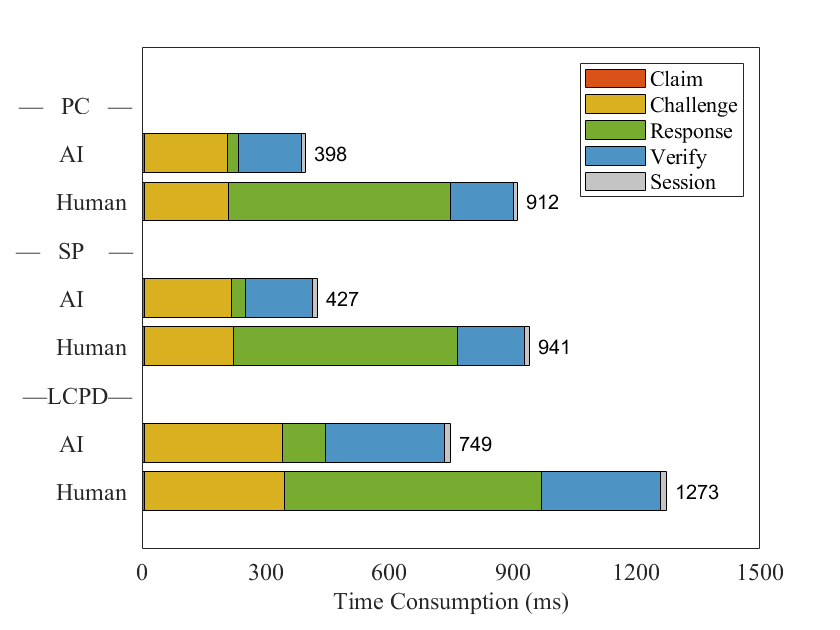}    }    \quad
    \subfigure[Virtual-to-physical tracing]{
    		\label{fig:TrackTime}    \includegraphics[width=8.3cm]{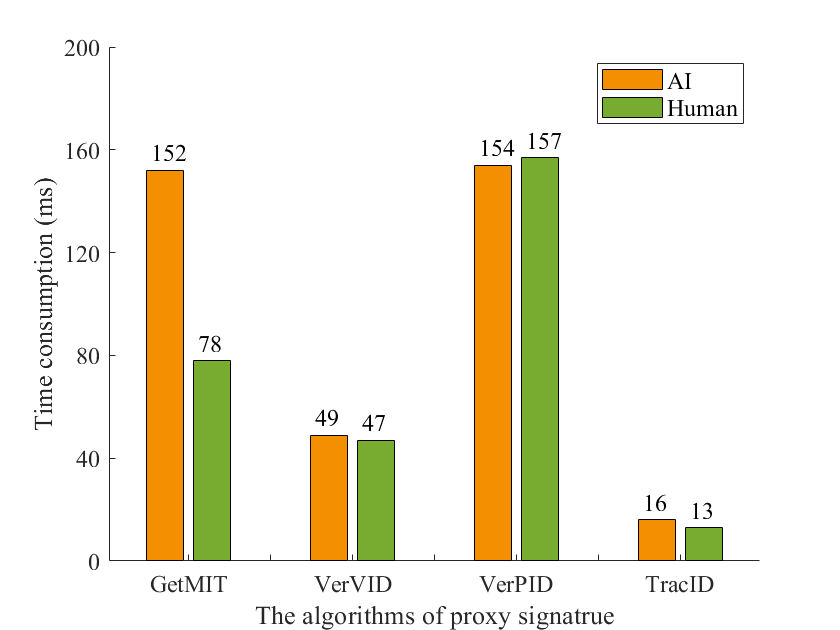}    }    \quad
    \caption{The time consumption on different types of operations.}

\end{figure*}

\subsection{Consumption  of CPS}

The performance of the proposed signature scheme CPS directly affects the authentication protocols. In this part, we analyze the communication overhead and computational cost of CPS to provide a basis for evaluating the performance of the proposed protocols.

\textbf{Communication Overhead of CPS:} To analyze the communication overhead of CPS, we compare the signature length of our scheme with that of similar schemes \cite{Verma2019CBPS,Verma2020CBPS,Qiao2022ProxTII,Yang2020MBP} as shown in the TABLE \ref{tab:CommOver}. It can be seen that the bit length of the proxy signature is $1\,|G|$ in both Verma’s scheme \cite{Verma2019CBPS} and ours scheme. However, the proxy signer in Verma's scheme only generates proxy signatures for a designated verifier, which is not suitable for AI-driven avatars. Therefore, the proposed signature scheme supports AI-driven avatars to generate proxy signatures with shorter length for arbitrary verifiers, reducing the communication overhead of authentication system.

\textbf{Computational Cost of CPS:} Since the CPS is inspired by Verma's scheme\cite{Verma2019CBPS} and Qiao's scheme\cite{Qiao2022ProxTII}, we compare the computational cost with these two schemes. In our scheme, $PVer$ is equal to $DVer$. The computational cost is shown as the TABLE \ref{tab:ComputationCost}, in which the computational cost of our scheme is slightly higher than that of the other two schemes. But for $PSig$, its computational cost is the same as that of the other two schemes. Thus, for a proxy signer in our scheme, he/she generates the proxy signatures at the same computational cost as Verma's and Qiao's schemes.

\textbf{Time Consumption of CPS:} Qiao's scheme \cite{Qiao2022ProxTII} is a  a pairing free construction and is able to efficiently generate signatures. To further evaluate the time consumption of CPS, we run this scheme and Qiao's scheme on PC platform. In the simulations, we set the  batch size of signatures to 10, 20, 30, and 40, respectively. It can be seen from the Fig. \ref{fig:ProxySign} that in our scheme the average time consumption for $DGen$, $DVer$, $PSig$, and $PVer$ is about 8ms, 28ms, 4ms, and 28ms, respectively, which slightly higher than that of Qiao's scheme. But for $PSig$, the time consumption is the same as that of Qiao. Therefore, our scheme supports AI-driven avatars to generate proxy signatures in an efficient way.

\subsection{Consumption of  Authentication Protocols}

\textbf{Time Consumption of Login Authentication:} To evaluate the performance of the login authentication protocol, we analyze the time consumption on the key steps and the full steps. In the login protocol, we take algorithms $DGen$, $PVer$, and $PSig$ as key steps about the phases of submit, challenge, response, and verify. In the simulations, the user $A$ utilizes the devices PC, SP, and LCPD for authentication, respectively, while the server $S$ utilizes the same device PC for authentication. The time consumption is shown in Fig.\ref{fig:LoginAuth}. In terms of key steps, its consumption is within 300ms on the three device platforms, while the consumption on full steps is about 1000ms, meeting the actual application needs. Thus, the designed login protocol supports users to log into the metaverse through different device platforms.

\textbf{Time Consumption of Delegated Authentication:} To evaluate the performance of delegated authentication protocol, we utilize PC as the main device for all participants to execute this protocol. In the simulations, we set the delegation batch to 10, 20, 30, and 40, respectively. The time consumption is shown as the Fig.\ref{fig:DlegAuth}, in which the average time consumption of user, server, and AI proxy is about 600ms, 200ms, and 300ms, respectively. In general, a single delegation takes about 1100ms, which does not affect the interaction experience. Therefore, the delegated authentication protocol can be applied to the metaverse to transfer human users' signing ability.

\textbf{Time Consumption of Mutual Authentication:} To evaluate the performance of mutual authentication protocol, we set the avatar $A$ to be driven by an AI proxy and the avatar $B$ to be driven by a human user. In the simulations,  both parties utilize the same device to verify each other. The time consumption of mutual authentication is shown as the Fig.\ref{fig:MutAuthen}. What can be seen from the figure is that the time consumption of verifying AI-driven avatars on PC, SP, and LCPD platforms is about 400ms, 430ms, and 750ms, respectively. But for human-driven avatars, the time consumption on the three platforms are about 1000ms, in which extracting iris feature takes about 500ms.  Therefore, in practical applications, we recommend building dedicated hardware to optimize extracting iris feature or using background authentication to reduce latency.

\textbf{Time Consumption of Tracing Avatars:} The tracing process mainly involves four steps: get $MIT$ from blockchain, verify $VID$, verify $PID$, and track down the user's $ID$. The time consumption is shown as the Fig.\ref{fig:TrackTime}. In terms of getting $MIT$, the time consumption on AI-driven avatars is slightly higher than that of human-driven avatars. The reason is that AI-driven avatars involve two $MIT$s, namely the original manipulator’s $MIT$ and the AI proxy’s $MIT$, while human-driven avatars involve only one $MIT$, namely the user’s own $MIT$. In general, the time consumption of tracing an AI-driven avatar is roughly the same as that of tracing a human-driven avatar, where the total consumption for both types of avatars is less than 500ms. Therefore, the tracing method achieves virtual-to-physical tracing in an efficient way and helps to mitigate impersonation attacks by AI-driven avatars.

\section{Discussion}\label{sec:Discu}

Impersonation is a common attack in metaverse, while tracing avatars is one of the effective ways to mitigate such an attack. In the following, we further discuss the impersonation attack and the traceability, then show the limitations of the proposed solution, and finally think about the future work.

\textbf{Impersonation.} Metaverse allows users to create arbitrary appearance as their avatars. With the help of AI, attackers are able to construct AI-driven avatars to impersonate other users. Therefore, the impersonation attack by AI-driven avatars is an issue that the metaverse industry must solve. Since AI-driven avatars have just entered our lives, there is a lack of method to defend against such an attack. To this end, this work designs a traceable authentication method to trace an AI-driven avatar back to its original manipulator.

\textbf{Traceability. } Avatar's traceability is a key to mitigating metaverse crimes. This work introduces the manipulator's biometric feature into the identity factor of AI-driven avatars, such that both human-driven and AI-driven avatars are associated with their physical manipulator. To prevent honest users from being falsely accused, we propose a proxy signature scheme to ensure that malicious  reporters fail to forge the required identity parameters even if they can extensively interact with the human-driven and the AI-driven avatars of a target user. Based on the proposed signature scheme, we design a mutual authentication protocol to guarantee that victims can trace a human-driven avatar or an AI-driven avatar back to its original manipulator based on the avatar identity parameters.

\textbf{Limitations.} Our authentication method guarantees the traceability including both the human-driven and AI-driven avatars, forcing attackers to consider being traced and abandon impersonation attacks. The limitation of this method is that it cannot detect the abnormal identity about AI-driven avatars, and thus cannot completely avoid impersonation attacks during the interaction process. Therefore, it is necessary to further design an authentication method for AI-driven avatars to detect their abnormal identity. In addition, since we introduce personal features into AI-driven avatars, this may conflict with some practical applications, such as an avatar representing only a certain organization. Therefore, it is necessary to further consider the identity factors of AI-driven avatars and design corresponding authentication methods.

\textbf{Future Work.} In view of the above limitations, the future work needs to introduce new ideas such as machine learning \cite{Zhang2024VML} to construct an authentication method for avatars, ensuring that the abnormal identity of AI-driven avatars can be discovered in the interaction process. With this new authentication method,  users are able to identify impersonation attacks without subsequent tracing. In addition, it is necessary to build new identity models to endow AI-driven avatars with group attributes, supporting mutual authentication between individual avatars and group avatars.

\section{Conclusion} \label{sec:Conclu}
In metaverse, the impersonation by AI-driven avatars is a common attack. To mitigate this attack, this paper proposed a traceable authentication framework for human-driven and AI-driven avatars. This framework combines the original manipulator's iris feature and the AI proxy's public key to construct a user's identity model, ensuring that an AI-driven avatar is associated with its original manipulator. To transfer the manipulator's signing ability to an AI proxy, we proposed a chameleon proxy scheme signature. Based on the identity model and the signature scheme, we designed three authentication protocols to achieve user login, avatar delegation, and mutual authentication, while guaranteeing the virtual-to-physical traceability and defending against false accusation. All the above operations are completed in about 1s, meeting the needs of metaverse applications. We wish that the proposed authentication method for AI-driven avatars could bring a little reference to researchers in related fields.

\section*{Acknowledgment}  
This work is supported by the National Key Research and Development Program of China under Grant 2021YFB3101100; National Natural Science Foundation of China under Grant No. 62303126, No. 62362008,  No.62262058, No. 62272123;  Guizhou Provincial Science and Technology Projects No. ZK[2022]149; Project of High-level Innovative Talents of Guizhou Province under Grant [2020]6008; Science and Technology Program of Guizhou Province under Grant [2020]5017, [2022]065; Guizhou Provincial Research Project (Youth) for Universities under grant [2022]104.

\vspace{12pt}

\clearpage

\appendices

\begin{figure*}[htbp]
    \centering
     \subfigure[The attack game of forging original signature]{
     		\label{fig:OS_EUF_CMA}    \includegraphics[width=8.0cm]{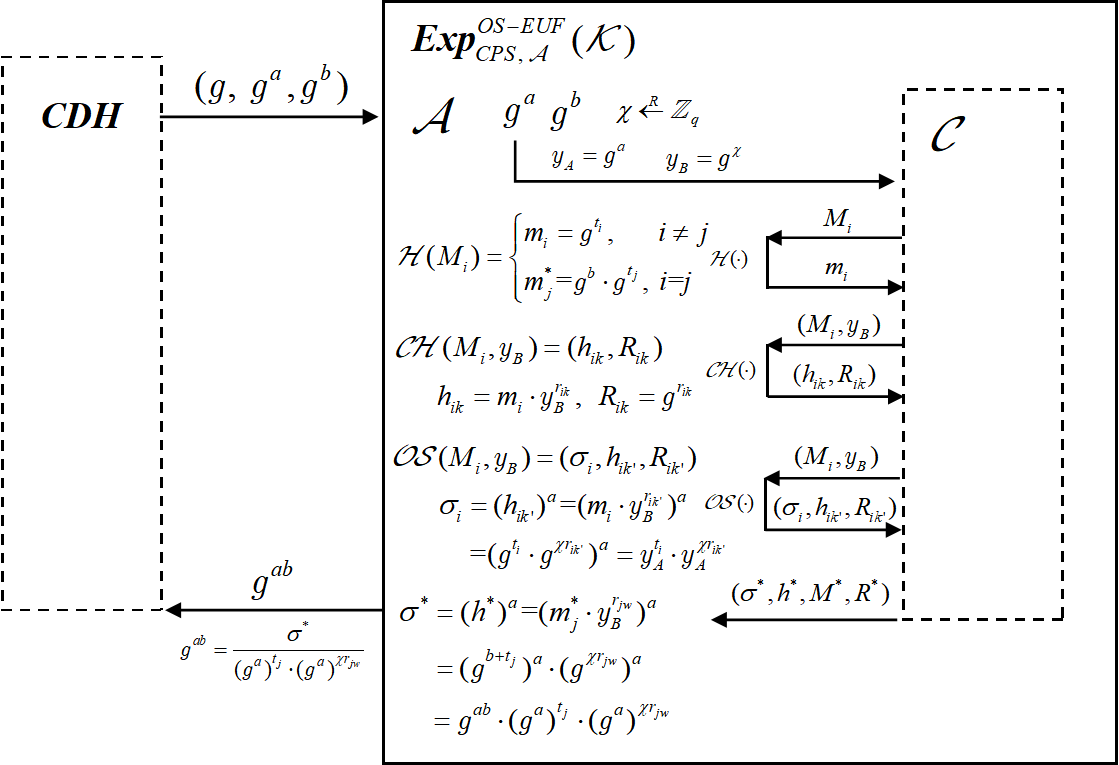}    }    \quad
    \subfigure[The attack game of forging proxy signature]{
    		\label{fig:PS_EUF_CMA}    \includegraphics[width=8.0cm]{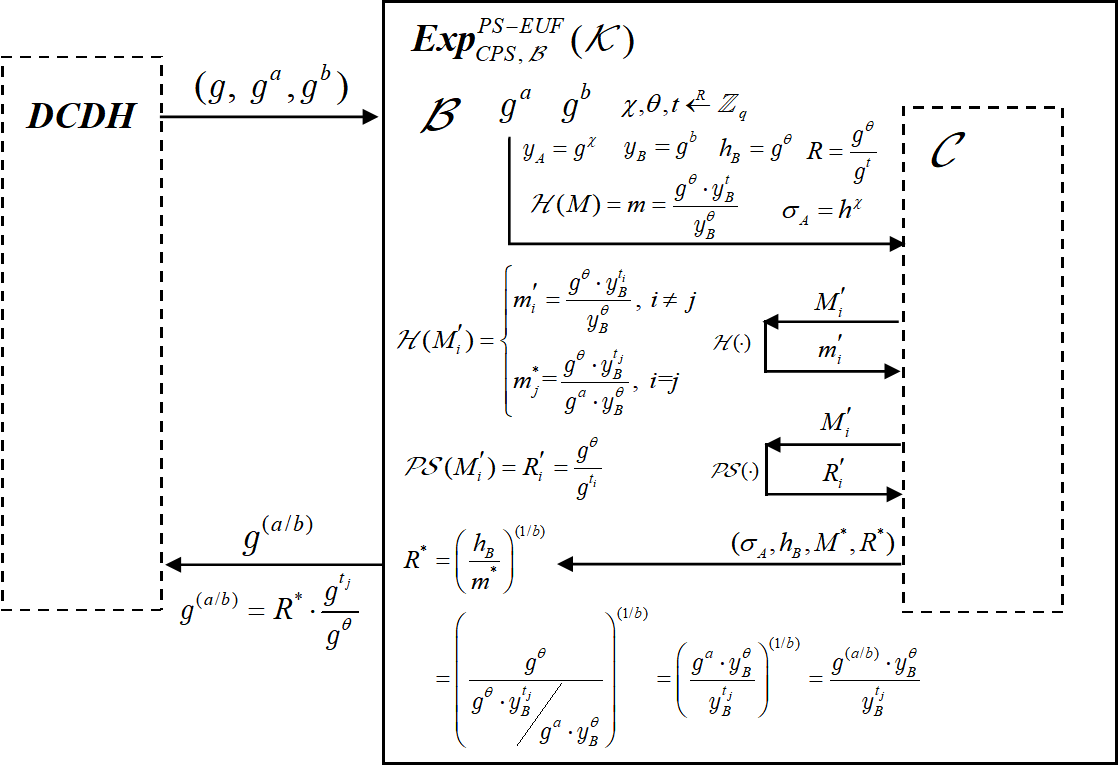}    }    \quad
    \caption{The attack games of the proposed signature scheme.}
\end{figure*}

\section{Security Proof for CPS} \label{sec:SecuProof}

In this part, we proves the security of the chameleon proxy signature scheme, including the OS-EUF-CMA under the CDH assumption and the PS-EUF-CMA under the DCDH assumption.

\subsection{OS-EUF-CMA} \label{sec:ProofOsEuf}

Let us review the \textbf{Theorem 1}. If the CDH assumption holds on $\mathbb{G}$, the chameleon proxy signature is OS-EUF-CMA.

\textbf{Proof 1.} Suppose that there is an adversary $\mathcal{C}$ who forges the original signature with a non-negligible advantage $\epsilon$, and there is an adversary $\mathcal{A}$ who is given a random instance  $(g, g^a, g^b) \in \mathbb{G}^3$ of the CDH problem. The purpose of $\mathcal{A}$  is to run $\mathcal{C}$ as a subroutine to ﬁnd the solution $g^{ab}$. For this purpose, $\mathcal{A}$ sets up a simulated attack game $Exp_{CPS,\mathcal{A}}^{OS\text{-}EUF}(\mathcal{K})$  about original signatures as shown in Fig.\ref{fig:OS_EUF_CMA}. The game is divided into three phases including initialization, inquiry, and forgery. The details are as follows:

(1) Initialization. $\mathcal{A}$ initializes the game environment through these four steps: (i) $\mathcal{A}$ sets the public keys $pk_A=y_A=g^a$ and $pk_B=y_B=g^\chi$ for $\mathcal{C}$, where $\chi \stackrel{R}{\leftarrow} \mathbb{Z}_q$. (ii) $\mathcal{A}$ constructs the oracles $\mathcal{H}(\cdot)$,  $\mathcal{CH}(\cdot)$, and  $\mathcal{OS}(\cdot)$ to simulate the algorithms $H(\cdot)$, $Hash(\cdot)$, and $DGen(\cdot)$, respectively; (iii) $\mathcal{A}$ constructs the lists $L_{\mathcal{H}}(\cdot)$, $L_{\mathcal{CH}}(\cdot)$, and $L_{\mathcal{OS}}(\cdot)$ to record the parameters about inquiring  $\mathcal{H}(\cdot)$,  $\mathcal{CH}(\cdot)$, and  $\mathcal{OS}(\cdot)$, respectively, where $L_\mathcal{H}=(M,t,m)$, $L_{\mathcal{CH}}=( M, y_B, r, h, R)$, and $L_{\mathcal{OS}}=(M, \sigma, h,R)$; (iv) $\mathcal{A}$ randomly selects $j, w \stackrel{R}{\leftarrow}\{1,2,\cdot\cdot\cdot,q_H\}$ as the hypothetical indexes of the forged signature, indicating that the $j$-th query of $\mathcal{H}(\cdot)$ and the $w$-th query of $\mathcal{CH}(\cdot)$ from $\mathcal{C}$ corresponds to the forged signature $(\sigma^*,h^*,M^*,R^*)$.

(2) Inquiry.  $\mathcal{C}$ inquires the oracles $\mathcal{H}(\cdot),\mathcal{CH}(\cdot)$, and $\mathcal{OS}(\cdot)$ at most $q_H$ times, while $\mathcal{A}$ responses the corresponding result as follow:

——$\mathcal{H}(M_i) \rightarrow m_i$. $\mathcal{A}$ retrieves $L_\mathcal{H}(\cdot)$ by $M_i$ and returns $m_i \leftarrow L_\mathcal{H}(M_i)$ upon receiving a query on $\mathcal{H}(\cdot)$ with $M_i$. If there is no corresponding entry, $\mathcal{A}$ randomly selects $t_i \stackrel{R}{\leftarrow} \mathbb{Z}_q$, calculates $m_i = g^{t_i}$, and adds a new entry $(M_i,t_i,m_i)$ to $L_\mathcal{H}(\cdot)$, where $i\in[1,...,q_H]$. It is worth noting that if $i=j$, $\mathcal{A}$ calculates $ m_{i=j}=m_j^*=g^b\cdot g^{t_j}$, which is used to construct the solution for the CDH problem.

——$\mathcal{CH}(M_i,y_B)\rightarrow (h_{ik},R_{ik})$. To generate a chameleon tuple $(h_{ik}, R_{ik})$ for $\mathcal{C}$, $\mathcal{A}$ retrieves $m_i \leftarrow L_\mathcal{H}(M_i)$, selects $r_{ik}\stackrel{R}{\leftarrow} \mathbb{Z}_q$, calculates the $(h_{ik}=m_i\cdot y_B^{r_{ik}}$ , $R_{ik}=g^{r_{ik}})$ for $\mathcal{C}$, and adds a new entry $( M_i, y_B, r_{ik}, h_{ik}, R_{ik})$ to $L_{\mathcal{CH}}(\cdot)$, where $k\in[1,...,q_H]$.

——$\mathcal{OS}(M_i, y_B) \rightarrow (\sigma_i,h_{ik^\prime},R_{ik^\prime})$. To generate an original signature $(\sigma_i,h_{ik^\prime},R_{ik^\prime})$ for $\mathcal{C}$, $\mathcal{A}$ retrieves $(t_i,m_i)$ $\leftarrow$ $ L_\mathcal{H}(M_i)$ and $(r_{ik^\prime},h_{ik^\prime},R_{ik^\prime}) \stackrel{R}{\leftarrow}  \{(r_{ik},h_{ik},R_{ik})\}_{k=1,...,q_H} \leftarrow L_{\mathcal{CH}}(M_i)$, where ${k^\prime} \neq w$. With these parameters, $\mathcal{A}$ calculates $\sigma_i={y_A^{t_i}\cdot y_A^{\chi r_{ik^\prime}}}$ for $\mathcal{C}$ and adds a new entry  $(M_i, \sigma_i, h_{ik^\prime},R_{ik^\prime})$ to $L_{\mathcal{OS}(\cdot)}$, in which
\begin{equation}
    \begin{split} \notag
        \sigma_i &=h_{ik^\prime}^a \\
        &=(m_i\cdot y_B^{r_{ik^\prime}})^a\\
        &=(g^{t_i}\cdot g^{\chi r_{ik^\prime}})^a\\
        &={y_A^{t_i}\cdot y_A^{\chi r_{ik^\prime}}}.\\
    \end{split}
\end{equation}

It is worth noting that if $i=j$, $\mathcal{A}$  stops and returns ``$\perp$'' indicating an error.

(3) Forgery. $\mathcal{C}$ outputs a forged signature $(\sigma^*$, $h^*$, $M^*$, $R^*)$. If $PVer(y_A,\sigma^*, h^*, M^*, R^*,y_B)= 1$ and $(M^*$, $\sigma^*$, $h^*$, $R^*) \notin L_{\mathcal{OS}}$, then $\mathcal{C}$ successfully forges a original signature and wins the game.

In the simulation $Exp_{CPS,\mathcal{A}}^{OS\text{-}EUF}(\mathcal{K})$, if the guess $j$ and $w$ from $\mathcal{A}$ are correct and $\mathcal{C}$ outputs a correct forgery,  $\mathcal{A}$ can be sure that 
\begin{equation}\nonumber
\begin{split}
    \sigma^* &= (h^*)^a=(m_j^*\cdot y_B^{r_{jw}})^a\\
    &=(g^{b+t_j})^a \cdot (g^{\chi r_{jw}})^a\\
    &=g^{ab}\cdot (g^a)^{t_j} \cdot (g^a)^{\chi r_{jw}}.\\
\end{split}
\end{equation}

Therefore, $\mathcal{A}$ is able to output the solution $g^{ab}$ of the CDH problem based on $\sigma^*$ as
\begin{equation}\nonumber
\begin{split}
    g^{ab}&=\frac{\sigma^*}{(g^a)^{t_j}\cdot (g^a)^{\chi r_{jw}}}\\
\end{split}
\end{equation}

The successful output $g^{ab}$ from $\mathcal{A}$ is determined by the following three events:

$\mathcal{E}_1$: No interruption is encountered during the interaction between $\mathcal{A}$ and $\mathcal{C}$.

$\mathcal{E}_2$: $\mathcal{C}$ produces a valid forgery $(\sigma^*,h^*,M^*, R^*)$.

$\mathcal{E}_3$: $\mathcal{E}_2$ occurs and the subscript of $M_i^*$ is $i=j$ and $R_k^*$ is $k=w$. Then

$Pr[\mathcal{E}_1]=(1-\frac{1}{q_H})^{q_H}$,

$Pr[\mathcal{E}_2|\mathcal{E}_1]=\epsilon$,

$Pr[\mathcal{E}_3|\mathcal{E}_1\mathcal{E}_2] =Pr[(i=j,k=w)|\mathcal{E}_1\mathcal{E}_2] =\frac{1}{q_H} \cdot \frac{1}{q_H}$ ,

Therefore, the advantage of $\mathcal{A}$ is
\begin{align}
\notag
Pr[\mathcal{E}_1\mathcal{E}_3]&=Pr[\mathcal{E}_1] \cdot Pr[\mathcal{E}_2|\mathcal{E}_1]\cdot Pr[\mathcal{E}_3|\mathcal{E}_1\mathcal{E}_2]\\
\notag
&=(1-\frac{1}{q_H})^{q_H}\cdot \frac{1}{q_H^2}\cdot \epsilon\\
\notag
&\approx \frac{\epsilon}{e\cdot q_H^2}.
\end{align}

Since the CDH assumption holds on $\mathbb{G}$, the advantage $\epsilon^\prime \approx \frac{\epsilon}{e\cdot q_H^2}$ of polytime adversary $\mathcal{A}$ is negligible. Therefore, the proposed signature scheme is OS-EUF-CMA. $\; \square$

\subsection{PS-EUF-CMA} \label{sec:ProofPsEuf}

Let us review the \textbf{Theorem 2}. If the DCDH assumption holds on $\mathbb{G}$, the chameleon proxy signature is PS-EUF-CMA.

\textbf{Proof 2.} Suppose that there is an adversary $\mathcal{C}$ who forges the proxy signature with a non-negligible advantage $\epsilon$, and there is an adversary $\mathcal{B}$ who is given a random instance  $(g, g^a, g^b) \in \mathbb{G}^3$ of the DCDH problem. The purpose of $\mathcal{B}$  is to run $\mathcal{C}$ as a subroutine to ﬁnd the solution $g^{(a/b)}$. For this purpose, $\mathcal{B}$ sets up a simulated attack game $Exp_{CPS,\mathcal{B}}^{PS\text{-}EUF}(\mathcal{K})$  about proxy signatures as shown in Fig.\ref{fig:PS_EUF_CMA}. The game is divided into three phases including initialization, inquiry, and forgery. The details are as follows:

(1) Initialization. $\mathcal{B}$ initializes the game environment through these six steps: (i) $\mathcal{B}$ sets the public keys $pk_A=y_A=g^\chi$ and $pk_B=y_B=g^b$ for $\mathcal{C}$, where $\chi \stackrel{R}{\leftarrow} \mathbb{Z}_q$;  (ii) $\mathcal{B}$ constructs the oracles $\mathcal{H}(\cdot)$ and  $\mathcal{PS}(\cdot)$ to simulate the algorithm $H(\cdot)$ and $PSig(\cdot)$, respectively; (iii) $\mathcal{B}$ constructs the lists $L_{\mathcal{H}}(\cdot)$ and $L_{\mathcal{PS}}(\cdot)$ to record the parameters about inquiring  $\mathcal{H}(\cdot)$ and  $\mathcal{PS}(\cdot)$, respectively, where $L_\mathcal{H}=(M,t,m)$ and $L_{\mathcal{PS}}=(M,R)$;  (iv) $\mathcal{B}$ selects $\theta,t \stackrel{R}{\leftarrow} \mathbb{Z}_q$, sets $h_B=g^\theta$, $R=\frac{g^\theta}{g^t}$,  $\mathcal{H}(M)=m=\frac{g^\theta \cdot y_B^t}{y_B^\theta}$, and calculates $\sigma_A=h^\chi$, by which $\mathcal{B}$ constructs the original signature $(\sigma_A,h,M,R)$; (v) $\mathcal{B}$ randomly selects $j\stackrel{R}{\leftarrow}\{1,2,\cdot\cdot\cdot,q_H\}$ as the hypothetical indexe of the forged signature, indicating that the $j$-th query of $\mathcal{H}(\cdot)$ from $\mathcal{C}$ corresponds to the forged proxy signature $(\sigma_A,h,M^*,R^*)$; (vi) $\mathcal{B}$ adds entries $(M,t,m)$, $(M,R)$ to $L_\mathcal{H},L_{\mathcal{PS}}$, respectively, and sends $(y_A, \sigma_A, h_B, M, R, y_B)$ to $\mathcal{A}$.

(2) Inquiry.  $\mathcal{C}$ inquires the oracles $\mathcal{H}(\cdot)$ and $\mathcal{PS}(\cdot)$ at most $q_H$ times, while $\mathcal{B}$ responses the corresponding result as follow:

——$\mathcal{H}(M_i^\prime) \rightarrow m_i^\prime$. $\mathcal{B}$ retrieves $L_\mathcal{H}(\cdot)$ by $M_i^\prime$ and returns $m_i^\prime \leftarrow L_\mathcal{H}(M_i^\prime)$ upon receiving a query on $\mathcal{H}(\cdot)$ with $M_i^\prime$. If there is no corresponding entry, $\mathcal{B}$ randomly selects $t_i \stackrel{R}{\leftarrow} \mathbb{Z}_q$, calculates $m_i^\prime = \frac{g^\theta \cdot y_B^{t_i}}{y_B^\theta}$, and adds a new entry $(M_i^\prime,t_i,m_i^\prime)$ to $L_\mathcal{H}$. It is worth noting that if $i=j$, $\mathcal{B}$ calculates $ m_{i=j}=m_j^*=\frac{g^\theta \cdot y_B^{t_j}}{g^a \cdot y_B^\theta}$, which is used to construct the solution for the DCDH problem.

——$\mathcal{PS}(M_i^\prime) \rightarrow R_i^\prime$.  $\mathcal{B}$ retrieves $L_\mathcal{PS}(\cdot)$ by $M_i^\prime$ and returns $R_i^\prime \leftarrow L_{\mathcal{PS}}(M_i^\prime)$  upon receiving a query on  $\mathcal{PS}(\cdot)$ with $M_i^\prime$. If there is no corresponding entry,  $\mathcal{B}$ retrieves $t_i\leftarrow L_\mathcal{H}(M_i^\prime)$  and calculates $R_i^\prime=\frac{g^\theta}{g^{t_i}}$. With this parameters, $\mathcal{B}$ adds a new entry  $(M_i^\prime, R_i^\prime)$ to $L_\mathcal{PS}$ and returns $R_i^\prime$ to $\mathcal{C}$. It is worth noting that if $i=j$, $\mathcal{B}$  stops and returns ``$\perp$'' indicating an error.

(3) Forgery. $\mathcal{C}$ outputs a forged signature $(\sigma_A$, $h_B$, $M^*$, $R^*)$. If $PVer(y_A,\sigma_A, h_B, M^*, R^*,y_B)= 1$ and $(M^*, R^*) \notin L_{\mathcal{PS}}$, then $\mathcal{C}$ successfully forges a proxy signature and wins the game.

In the simulation $Exp_{CPS,\mathcal{B}}^{PS\text{-}EUF}(\mathcal{K})$, if the guess $j$ from $\mathcal{B}$ is correct and $\mathcal{C}$ outputs a correct forgery, $\mathcal{B}$ can be sure that 
\begin{equation}\nonumber
\begin{split}
    R^* &= \Big(\frac{h_B}{m_j^*}\Big)^{(1/b)}\\
    &=\Big(\frac{g^\theta}{ (g^\theta \cdot y_B^{t_i}) / (g^a \cdot y_B^\theta)}\Big)^{(1/b)}\\
    &=\Big( \frac{g^a\cdot y_B^\theta}{y_B^{t_j}} \Big) ^{(1/b)}.\\
    &=\frac{g^{(a/b)}\cdot g^\theta}{g^{t_j}}\\
\end{split}
\end{equation}

Therefore, $\mathcal{B}$ is able to output the solution $g^{(a/b)}$ of the DCDH problem based on $R^*$ as
\begin{equation}\nonumber
\begin{split}
    g^{(a/b)}&= R^*\cdot \frac{g^{t_j}}{g^\theta}.\\
\end{split}
\end{equation}

The successful output $g^{(a/b)}$ from $\mathcal{B}$ is determined by the following three events:

$\mathcal{E}_1$: No interruption is encountered during the interaction between $\mathcal{B}$ and $\mathcal{C}$.

$\mathcal{E}_2$: $\mathcal{C}$ produces a valid forgery $(\sigma_A,h_B,M^*, R^*)$.

$\mathcal{E}_3$: $\mathcal{E}_2$ occurs and the subscript of $M_i^*$ is $i=j$. Then

$Pr[\mathcal{E}_1]=(1-\frac{1}{q_H})^{q_H}$,

$Pr[\mathcal{E}_2|\mathcal{E}_1]=\epsilon$,

$Pr[\mathcal{E}_3|\mathcal{E}_1\mathcal{E}_2] =Pr[i=j|\mathcal{E}_1\mathcal{E}_2] =\frac{1}{q_H}$ .

Therefore, the advantage of $\mathcal{B}$ is
\begin{align}
\notag
Pr[\mathcal{E}_1\mathcal{E}_3]&=Pr[\mathcal{E}_1] \cdot Pr[\mathcal{E}_2|\mathcal{E}_1]\cdot Pr[\mathcal{E}_3|\mathcal{E}_1\mathcal{E}_2]\\
\notag
&=(1-\frac{1}{q_H})^{q_H}\cdot \frac{1}{q_H}\cdot \epsilon(\mathcal{K})\\
\notag
&\approx \frac{\epsilon}{e\cdot q_H}.
\end{align}

Since the DCDH assumption holds on $\mathbb{G}$, the advantage $\epsilon^\prime \approx \frac{\epsilon}{e\cdot q_H}$ of polytime adversary $\mathcal{B}$ is negligible. Therefore, the proposed signature scheme is PS-EUF-CMA. $\; \square$

\section{Authentication Protocols for Avatars}\label{sec:DetailsProto}

The main goal of this work is to build a traceable authentication method for human-driven and AI-driven avatars, by which victims can trace malicious avatars to its original manipulator. In this part, we firstly introduce the login authentication protocol to create a human-driven avatar. Secondly, we introduce the delegated authentication protocol to ensure that a user can delegate his/her avatar to an AI proxy. Thirdly, we introduce the mutual authentication protocol to achieve verification between human-driven and AI-driven avatars while guaranteeing the virtual-to-physical traceability.

\subsection{Login Authentication Protocol} \label{sec:LoginProto}
The login authentication protocol assumes that a user $A$ has previously registered an empty avatar $Avatar_A=\{SN_A,Aid_A,\text{-},\text{-},\text{-},\text{-},\text{-}\}$ on the cloud server $S$, where the symbol ``-'' means that the parameter value is empty and needs further assignment. The authentication protocol is shown as Fig.\ref{fig:LoginProt}.

 \textbf{(1) Claim.} In the login claim, $A$ provides $S$ with the avatar's virtual identity $VID_A$ by the following steps: (i) $A$ generates $(\sigma_A$, $h_A, R_A) \leftarrow DGen(sk_A$, $M_A,pk_A)$ to form the original signature $(\sigma_A$, $h_A$, $M_A, R_A)$, where $M_A$ is rendered as the avatar's visual appearance and $pk_A$ means that $A$ delegates his/her signing ability to himself/herself; (ii) $A$ submits  $\{SN_A$, $Aid_A$, $\sigma_A$, $h_A$, $VID_A \}$ to $S$, where $VID_A=(M_A, R_A)$.

 \textbf{(2) Challenge.} $S$ verifies $VID_A$ and throws a random challenge to $A$ by the following steps: (i) $S$ gets $MIT_A$ from the blockchain according to $SN_A$; (ii) $S$ verifies IDP's signature on $MIT_A$ to obtain $pk_A$ from $MIT_A$; (iii) $S$ verifies $VID_A$ by $PVer(pk_A, \sigma_A, h_A, M_A, R_A, pk_A)$ to ensure the validity of $VID_A$; (iv) $S$ throws a random challenge $C_a$ to $A$.

\textbf{(3) Response.} $A$ responses the avatar's physical identity $PID_A$ by the following steps: (i) $A$ samples an iris feature $M_a^\prime$ to form $M_A^\prime=M_a^\prime||C_a$; (ii) $A$ generates $R_A^\prime \leftarrow PSig(sk_A,h_A,M_A^\prime)$ and responses $PID_A = (M_A^\prime,R_A^\prime)$.

\textbf{(4) Verify.} $S$ verifies $PID_A$ and creates an avatar $Avatar_A$ through the following steps : (i) $S$ parses the $\tilde{M}_a^\prime$ and $\tilde{C}_a$ from $M_A^\prime$ to check $\tilde{C}_a \stackrel{?}{=} C_a$, which determines the freshness of $M_A^\prime$; (ii) $S$ verifies the match between $\tilde{M}_a$ and $T_A$ in $MIT_A$; (iii) $S$ verifies the $PID_A$ by $PVer(pk_A, \sigma_A, h_A, M_A^\prime, R_A^\prime, pk_A)$; (iv)  $S$ creates $Avatar_A=(SN_A, Aid_A, SN_A, \sigma_A, h_A, VID_A, \text{-})$ for $A$, where the proxy's serial number $SN_P=SN_A$ indicates that the avatar is driven by the human user $A$.

\subsection{Delegated Authentication Protocol} \label{sec:DelegProto}

\begin{figure*}[htbp]
\begin{center}
    \includegraphics[width=0.97\textwidth]{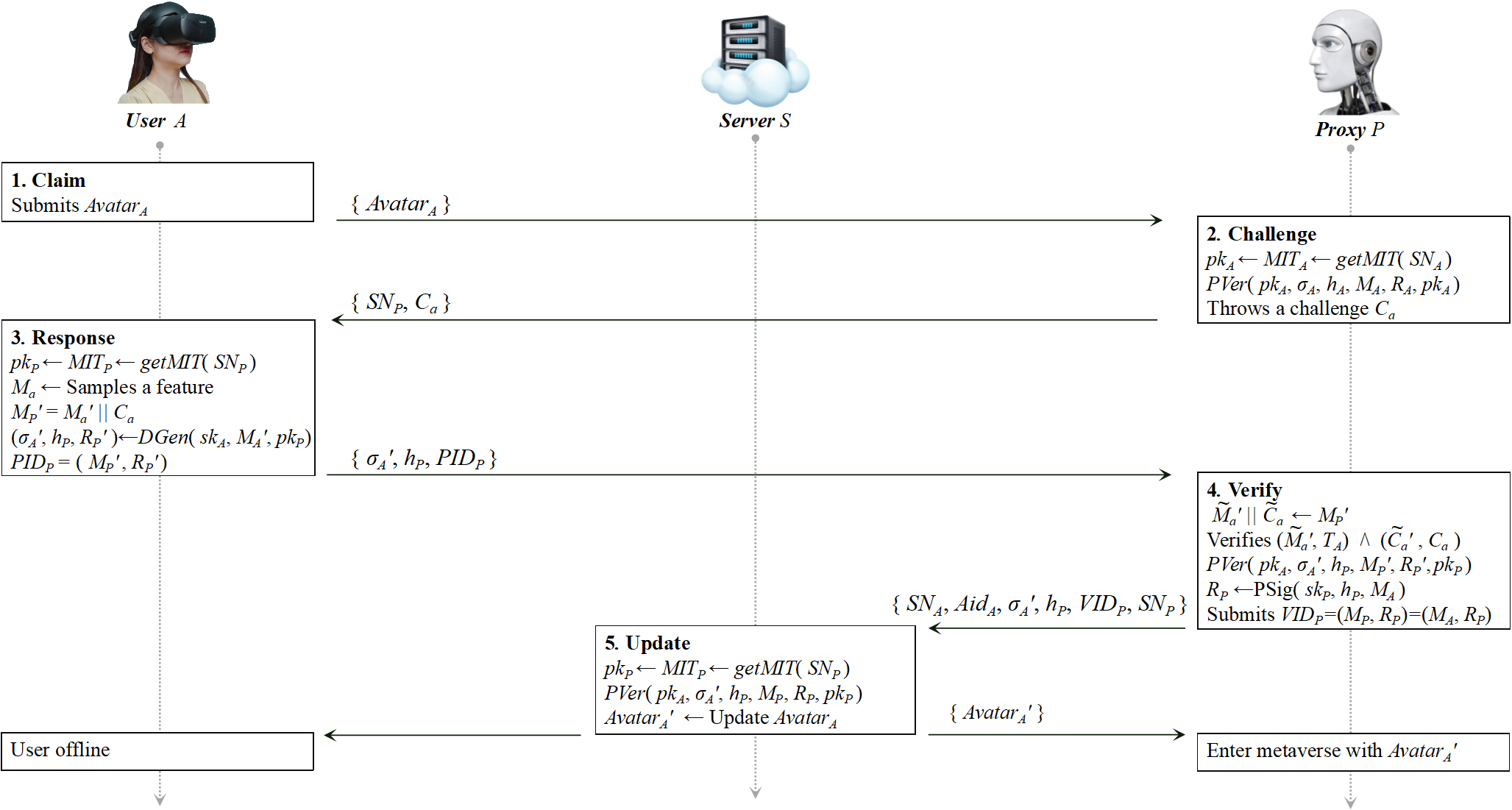}
    \caption{\small{Delegated authentication protocol.}}
    \label{fig:ProxyProt}
 \end{center}
\end{figure*}

The delegated authentication protocol enables a user $A$ to delegate his/her avatar $Avatar_A$ to an AI proxy $P$, which generates the corresponding AI-driven avatar $Avatar_A^\prime$. During the delegation process as shown in Fig.\ref{fig:ProxyProt}, $A$ generates an original signature to form the physical identity $PID_P$ of $Avatar_A^\prime$ based on $A$'s iris feature and $P$'s public key, while $P$ generates the corresponding proxy signature to form the virtual identity $VID_P$ of $Avatar_A^\prime$. Among them, $PID_P$ ensures that the $Avatar_A^\prime$ is associated with its original manipulator $A$. Details are as follows:  

\textbf{(1) Claim.} To delegate $Avatar_A$ to $P$, $A$ provides $P$ with the  $Avatar_A=(SN_A$, $Aid_A$, $SN_A$, $\sigma_A$, $h_A$, $VID_A, PID_A)$. 

\textbf{(2) Challenge.}
$P$ verifies $VID_A$ of $Avatar_A$ and throws challenge information $\{SN_P,C_a\}$ to $A$ by the following steps: (i) $P$ obtains $pk_A$ from $MIT_A$ based on the $SN_A$ in $Avatar_A$;  (ii) $P$ verifies the $VID_{A}=(M_A,R_A)$ by $PVer(pk_A,\sigma_A,h_A,M_A,R_A,pk_A)$; (iii) $P$ generates a random challenge $C_a$ and throws $\{SN_P,C_a\}$ to $A$, where $SN_P$ in $MIT_P$.

\textbf{(3) Response.}
$A$ returns the parameters $\{\sigma_A^\prime, h_P, PID_P \}$ as a response by the following steps: (i) $A$ obtains $P$'s public key $pk_P$ in $MIT_P$ according to $SN_P$; (ii) $A$ samples an iris feature $M_a^\prime$ to construct $M_P^\prime=M_a^\prime||C_a$ ; (iii) $A$ generates an original signature  $(\sigma_A^\prime, h_P, R_P^\prime) \leftarrow DGen(sk_A,M_A^\prime,pk_P)$ to construct $PID_P=(M_A^\prime,R_P^\prime)$.

\textbf{(4) Verify.}
$P$ submits the identity parameters of $\{SN_A,Aid_A,\sigma_A^\prime,h_P,VID_P,SN_P\}$ to $S$, indicating that $P$ has the signing ability of $Aid_A$. The specific steps are as follows: (i) $P$ verifies $PID_P$ and generates the proxy signature $R_P\leftarrow PSig(sk_P,h_P,M_A) $; (ii) $P$ constructs the avatar's virtual identity $VID_P=(M_P,R_P)=(M_A,R_P)$, where $M_A$ is taken from $VID_A=(M_A,R_A)$ in the human-driven avatar $Avatar_A$.

\textbf{(5) Transfer.}
$S$ transfers $Avatar_A$'s driving ability from $A$ to $P$ as follows: (i) $S$ verifies $VID_P$ by $PVer(pk_A$, $\sigma_A^\prime$, $h_P$, $M_P$, $R_P$, $pk_P)$; (ii)  $S$ updates $Avatar_A$ to $Avatar_A^\prime=\{SN_A$, $Aid_A$, $SN_P$, $ \sigma_A^\prime, h_P, VID_{P}, \text{-}\}$; (iii) $S$ sends $Avatar_A^\prime$ to $P$ and forces the user $A$ offline.

\subsection{Mutual Authentication Protocol} \label{sec:MutualProto}

\begin{figure*}[htbp]
    \centering
     \subfigure[Login authentication protocol]{
     		\label{fig:LoginProt}    \includegraphics[width=0.46\textwidth]{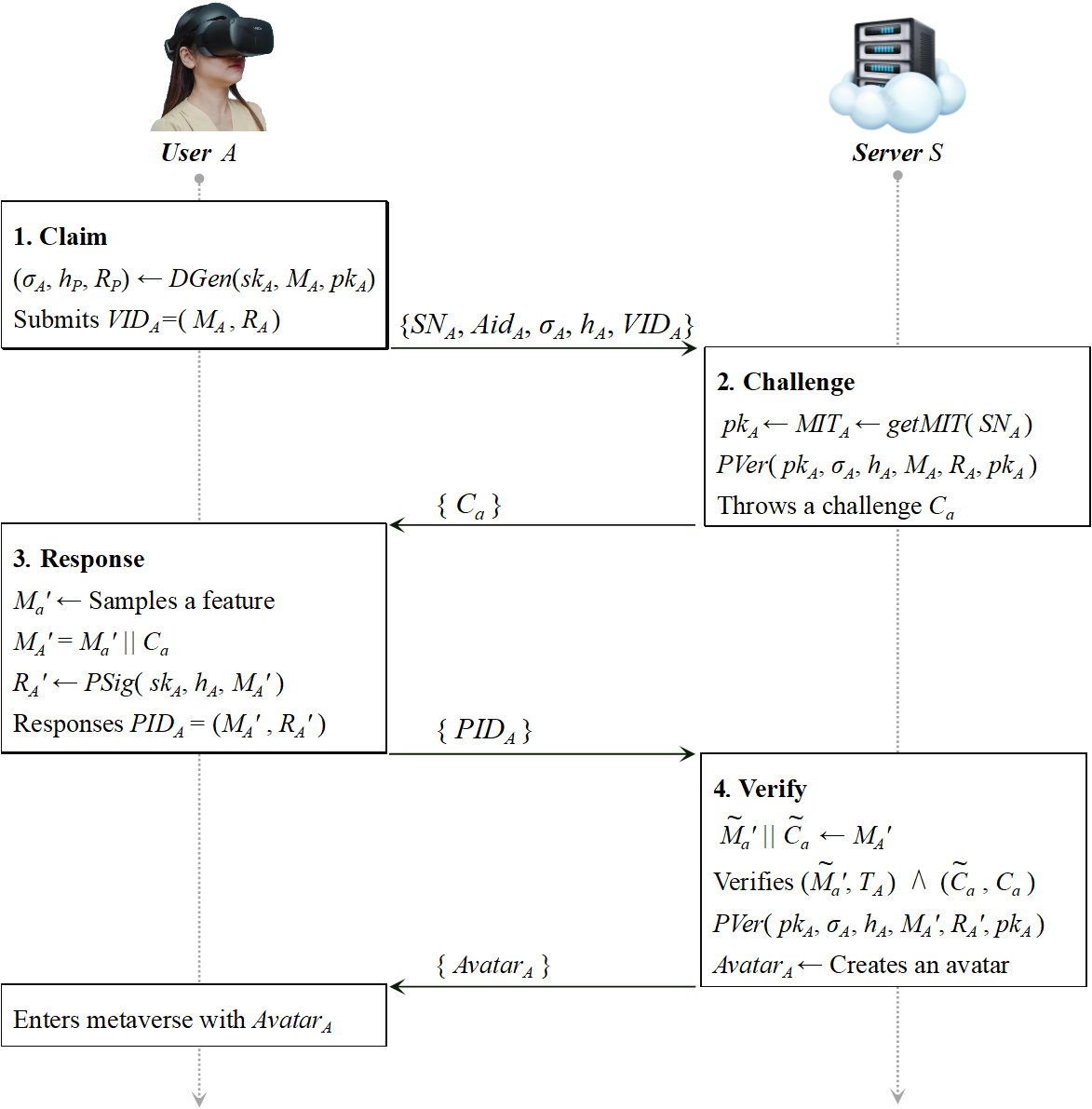}    }    \quad
    \subfigure[Mutual authentication protocol]{
    		\label{fig:MutualProt}    \includegraphics[width=0.46\textwidth]{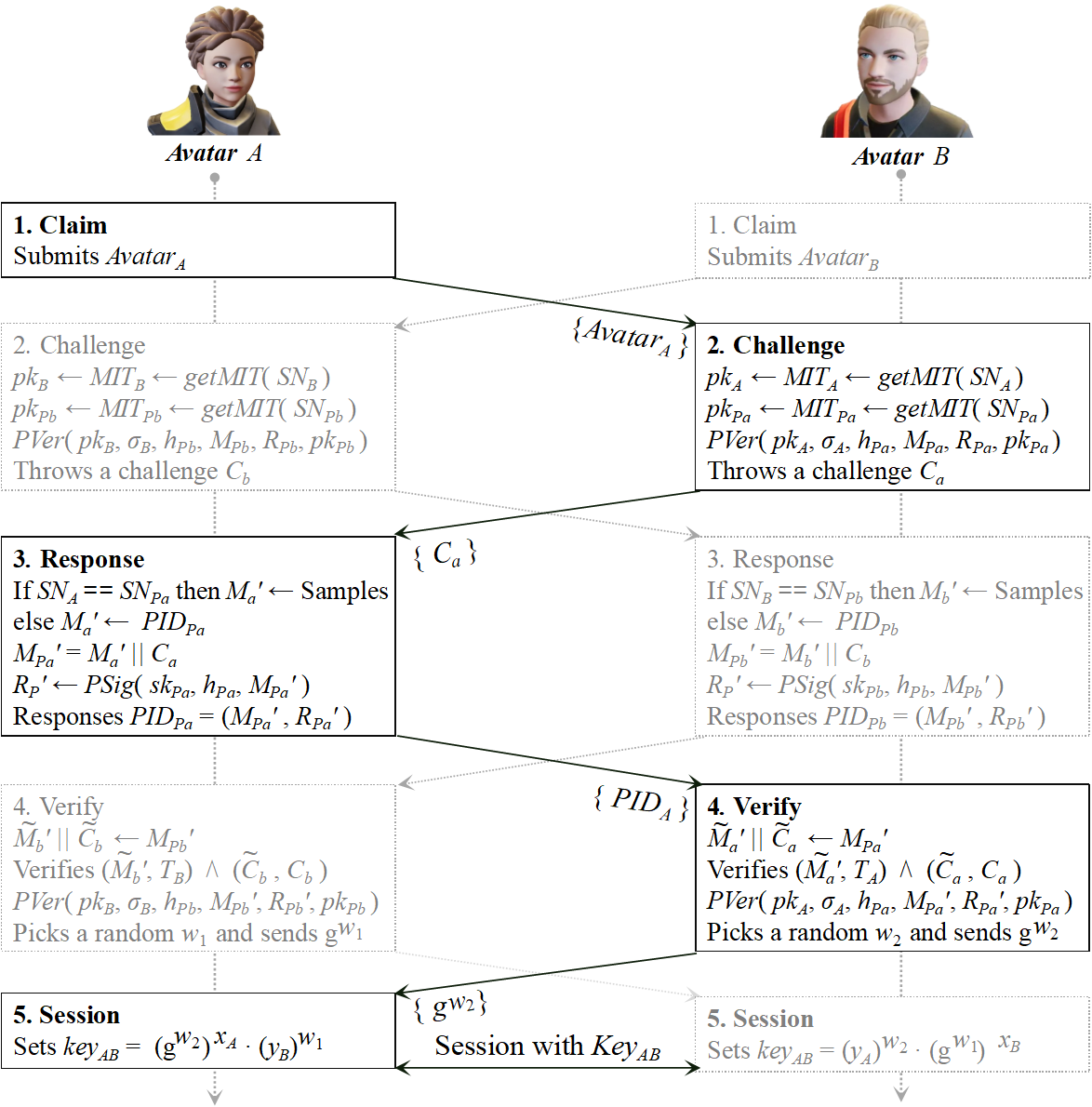}    }    \quad
    \caption{The login authentication protocol and the mutual authentication protocol.}
\end{figure*}

The mutual authentication protocol achieves verification between avatars  including both human-drive and AI-driven avatars to establish a session key. Its formal description is shown as Fig.\ref{fig:MutualProt}, in which both parties use the same steps to verify each other. It is worth noting that, in the phase of $Verify$, $A$ picks a random parameter $w_1$ and sends the element $g^{w_1} \in \mathbb{G}$ to $B$ , while $B$ picks a random parameter $w_2$ and sends the element $g^{w_2} \in \mathbb{G}$ to $A$. With these parameters, $A$ establishes the session key as $Key_{AB}=(g^{w_2})^{x_A}\cdot (y_B)^{w_1}$, while $B$ establishes the session key as $Key_{AB}= (y_A)^{w_2}\cdot(g^{w_1})^{x_B}$.  

Since the mutual authentication protocol is constructed by the login authentication protocol and the delegated authentication protocol, we do not provide details of  the mutual authentication protocol.

\end{document}